# Temperature Evolution of Magnon Propagation Length in $Tm_3Fe_5O_{12}$ Thin Films: Roles of Magnetic Anisotropy and Gilbert Damping

Amit Chanda[1], Christian Holzmann[2], Noah Schulz[1], Aladin Ullrich[2], Derick DeTellem[1], Manfred Albrecht[2*], Miela J. Gross[3], Caroline A. Ross[3*], Dario A. Arena[1], Manh-Huong Phan[1], and Hariharan Srikanth[1*]

[1]*Department of Physics, University of South Florida, Tampa, Florida 33620, USA*

[2]*Institute of Physics, University of Augsburg, 86159 Augsburg, Germany*

[3]*Department of Materials Science and Engineering, Massachusetts Institute of Technology, Cambridge, Massachusetts 02139, USA*

*Corresponding authors: manfred.albrecht@physik.uni-augsburg.de; caross@mit.edu; sharihar@usf.edu*

**ABSTRACT**

The magnon propagation length, $\langle \xi \rangle$ of a ferro-/ferrimagnet (FM) is one of the key factors that controls the generation and propagation of thermally-driven magnonic spin current in FM/heavy metal (HM) bilayer based spincaloritronic devices. For the development of a complete physical picture of thermally-driven magnon transport in FM/HM bilayers over a wide temperature range, it is of utmost importance to understand the respective roles of temperature-dependent Gilbert damping ($\alpha$) and effective magnetic anisotropy ($K_{eff}$) in controlling the temperature evolution of $\langle\xi\rangle$. Here, we report a comprehensive investigation of the temperature-dependent longitudinal spin Seebeck effect (LSSE), radio frequency transverse susceptibility, and broadband ferromagnetic resonance measurements on $Tm_3Fe_5O_{12}$ (TmIG)/Pt bilayers grown on different substrates. We observe a remarkable drop in the LSSE voltage below 200 K independent of TmIG film thickness and substrate choice. This is attributed to the noticeable increases in effective magnetic anisotropy field, $H_K^{eff}$ ($\propto K_{eff}$) and $\alpha$ that occur within the same temperature range. From the TmIG thickness dependence of the LSSE voltage, we determined the temperature dependence of $\langle\xi\rangle$ and highlighted its correlation with the temperature-dependent $H_K^{eff}$ and $\alpha$ in TmIG/Pt bilayers, which will be beneficial for the development of rare-earth iron garnet-based efficient spincaloritronic nanodevices.

## 1. INTRODUCTION

In recent years, interface-engineered bilayer thin films have gained intense attention of the materials science community because of their multifunctionality and emergent physical properties ranging from ferroelectricity[1] and magnetism[2] to spin-electronics[3]. Bilayers comprised of insulating rare-earth iron garnet (REIG) and heavy metal (HM) form the most appealing platform to generate, transmit, and detect pure spin currents in the field of spin-based-electronics[4–6]. The interplay of damping and magnon propagation length ($\langle \xi \rangle$) of the REIG layer and spin-orbit coupling (SOC) of the HM layer leads to a wide range of emergent spintronic phenomena in this fascinating class of heterostructures, including the spin Hall effect[7], spin-orbit torque[8,9], spin-pumping effect (SPE)[10], and the longitudinal spin Seebeck effect (LSSE)[11–13]. The discovery of the SSE[14] instigated a new generation of spintronic nanodevices facilitating electrical energy harvesting from renewable thermal energy wherein a magnonic spin current is thermally generated and electrically detected by applying a temperature gradient across a magnetic insulator (MI)/HM bilayer[15]. Unlike magnetostatic spin waves with millimeter-range propagation lengths, $\langle \xi \rangle$ for thermally generated magnons is significantly smaller, a few hundreds of nanometers[16]. In the framework of an atomistic spin model based on linear spin-wave theory, it was theoretically shown[17,18] that thermally generated magnons have a broad frequency ($f$) distribution with $f_{minimum} = 2K_{eff}/[h(1 + \alpha^2)]$ and $f_{maximum} = 4K_{eff}/[h(1 + \alpha^2)]$, where $h$ is the Planck constant, $K_{eff}$ is the effective magnetic anisotropy constant and $\alpha$ is the Gilbert damping parameter. While the high-$f$ magnons experience stronger damping, low-$f$ magnons possess a very low group velocity, and hence, the majority of the thermally generated magnons become damped on shorter length-scales[17,18]. Therefore, only the subthermal magnons, *i.e.*, the low-$f$ magnons dominate the long-range thermo-spin transport[19–21]. Within this hypothesis, it was predicted that $\langle\xi\rangle$ is inversely proportional to both $\alpha$ and $\sqrt{K_{eff}}$.[17,18]

$Y_3Fe_5O_{12}$ (YIG) has been a widely explored MI for generating and transmitting pure spin currents due to its ultra-low damping ($\alpha \approx 10^{-4}$-$10^{-5}$) and large $\langle\xi\rangle$ (~100-200 nm) [11,13,17]. This has led to a drastic increase in research over the last few decades, aimed at enhancing the spin current injection efficiency across the MI/HM interface by reducing the conductivity mismatch between the MI and HM layers by introducing atomically thin semiconducting interlayers[22–28] and enhancing the interfacial spin-mixing conductance.[29–31] Hariharan's group has explored the roles of bulk and surface magnetic anisotropies in LSSE in different REIG-based MI/HM bilayers[12,13], whereas a recent study highlights the influence of damping on SPE and LSSE in a compensated ferrimagnetic insulator.[32] It has also been demonstrated that the LSSE in YIG/Pt bilayers varies inversely with intrinsic Gilbert damping of the YIG films, however, the LSSE coefficient does not show any significant correlation with the enhanced damping due to SPE in YIG/Pt bilayers.[33] All these studies highlight the important roles of both magnetic anisotropy and Gilbert damping in thermally generated magnon propagation in MI/HM bilayers.

By investigating the YIG thickness dependence of the local LSSE measurements in YIG/Pt, Guo et al.[34] determined the temperature (*T*) dependence of $\langle\xi\rangle$ and found a scaling behavior of $\langle\xi\rangle \propto T^{-1}$. On the contrary, by employing non-local measurement geometries, Cornelissen et al. demonstrated that the magnon diffusion length for thermally driven magnonic spin currents ($\lambda_{th}^m$) of YIG decreases with decreasing temperature over a broad temperature range.[35] Gomez-Perez et al. reported similar observations and demonstrated that the temperature dependence of $\lambda_{th}^m$ is independent of the YIG thickness.[19] The different trends of the temperature dependent characteristic critical length scales for thermally generated magnon propagation in YIG observed by different groups indicates distinct temperature evolutions of $\alpha$ and $K_{eff}$ in the YIG films grown by these groups. In other words, different thin film growth conditions and sample dependent changes in the physical properties can give rise to different temperature dependences of both $\alpha$ and $K_{eff}$ and hence $\langle\xi\rangle$. For the development of a complete physical picture of LSSE in these REIGs over a wide temperature range, it is of utmost importance to comprehend the respective roles of both $\alpha$ and $K_{eff}$ simultaneously in determining the temperature evolution of $\langle\xi\rangle$, which remains largely unexplored.

Although YIG is considered as a benchmark system for LSSE,[11,34] there is only a limited number of studies that explore temperature dependent LSSE in other iron garnets.[12,32,36,37] For example, $Gd_3Fe_5O_{12}$ (GdIG) which is a ferrimagnetic insulator with magnetic compensation temperature ($T_{Comp}$) close to room temperature, shows a sign-inversion in the LSSE voltage[12] as well as in the spin-Hall anomalous Hall effect[38] around its magnetic compensation. However, the Gilbert damping in GdIG diverges over a broad temperature range around its $T_{Comp}$ which makes it difficult to probe the temperature evolution of $\alpha$ and its contribution towards $\langle\xi\rangle$ over a wide temperature range around the $T_{Comp}$.[32] Apart from YIG and GdIG, there has been a renaissance of research interest in another member of the REIG family: $Tm_3Fe_5O_{12}$ (TmIG) due to its wide-ranging extraordinary magnetic properties[6] *e.g.*, strain-tunable perpendicular magnetic anisotropy (PMA),[39] chiral and topological spin textures,[40] and interfacial Dzyaloshinskii-Moriya interaction[40,41] combined with low coercivity[6] which make this system a promising candidate for numerous efficient spintronic applications, such as spin-orbit torque induced magnetization switching,[8,42] current-induced domain-wall motion,[43] and spin Hall–topological Hall effects[44,45]. Recently, the LSSE has been investigated in TmIG/Pt bilayers with PMA at room temperature, and shown to exhibit high interfacial spin transparency and spin-to-charge conversion efficiency at the TmIG/Pt interface[46]. TmIG has a higher Gilbert damping parameter ($\approx 10^{-2}$)[6] compared to YIG, and unlike GdIG, TmIG does not exhibit any magnetic compensation in the temperature range between 1.5 and 300 K[47,48], which allows us to probe the relative contribution of $\alpha$ towards the temperature evolution of $\langle \xi \rangle$ and hence the LSSE over a broad temperature range close to the room temperature. However, the temperature evolution of LSSE and hence $\langle \xi \rangle$ as well as their relationship with $\alpha$ and $K_{eff}$ in TmIG/Pt bilayers are yet to be explored, which would be of critical importance for REIG-based efficient magnonic device applications. Here, we have performed a comprehensive investigation of the temperature-dependent LSSE, radio frequency (RF) transverse susceptibility (TS), and broadband ferromagnetic resonance (FMR) of TmIG/Pt bilayers grown on different substrates. From the TmIG thickness dependence of the LSSE voltage, we determined the temperature dependence of $\langle \xi \rangle$ and highlighted its correlation with the temperature-dependent effective magnetic anisotropy field, $H_K^{eff}$ ($\propto K_{eff}$) and $\alpha$ in TmIG/Pt bilayers.

## 2. RESULTS AND DISCUSSION

### 2. 1. Structural Characterization

Single-crystalline TmIG films with different thicknesses were grown on (111)-oriented $Gd_3Sc_2Ga_3O_{12}$ (GSGG) and $Gd_3Ga_5O_{12}$ (GGG) substrates by pulsed laser deposition (**see Methods**). The high crystalline quality of the TmIG films was confirmed by X-ray diffraction (XRD). **Figure 1**(a) shows the $\theta - 2\theta$ X-ray diffractograms of the GSGG/TmIG($t$) films with different TmIG film thickness $t$ ($t$ = 236, 150, 89, 73, 46 and 28 nm).

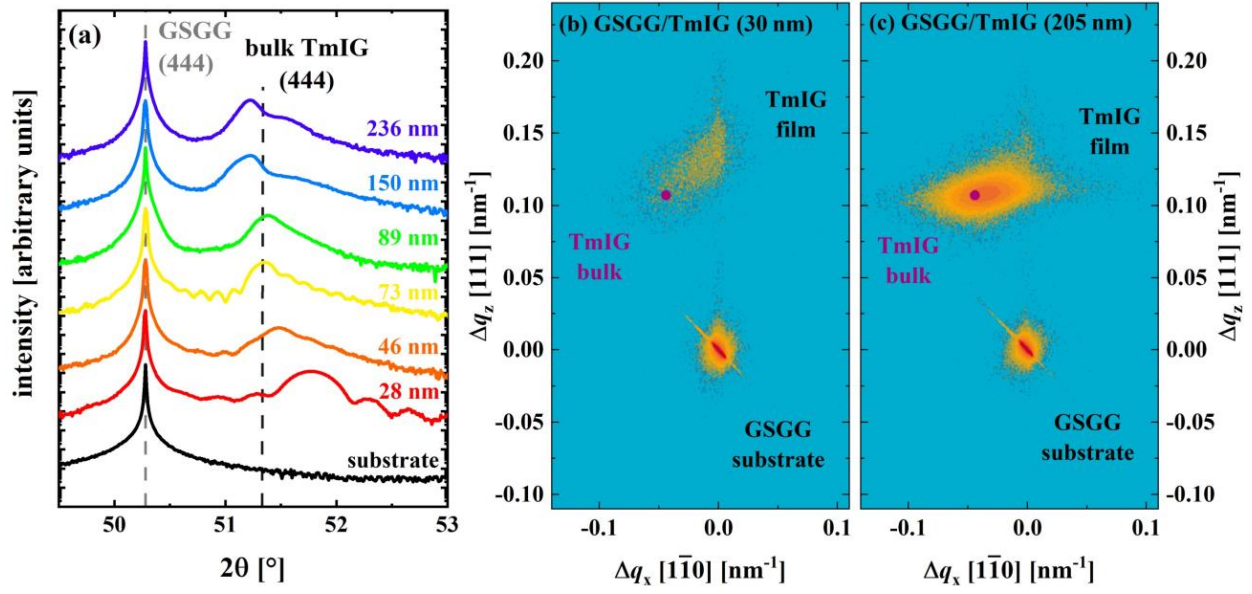

**Figure 1**. **Structural and Morphological characterization.** (a) $\theta - 2\theta$ X-ray diffractogram of the GSGG/TmIG($t$) films with different film thickness $t$ ($t$ = 236, 150, 89, 73, 46 and 28 nm). The reciprocal space maps recorded in the vicinity of the (642) reflection for **(b)** GSGG/TmIG(30 nm) and **(c)** GSGG/TmIG(205 nm) films. For the thinner film (30 nm), the TmIG film peak matches the IP lattice constant of the GSGG substrate, whereas for the thicker film (205nm), the TmIG film is largely relaxed.

The substrate choice and the TmIG film thickness influence the strain state of the film. **Figs. 1**(b) and (c) show the reciprocal space maps in the vicinity of the (642) reflection for the GSGG/TmIG (30 nm) and GSGG/TmIG (205 nm) films, respectively. For the thinner film (30 nm), the TmIG film $q_x$ matches the in-plane (IP) lattice spacing of the GSGG substrate indicating coherent growth, and the out-of-plane (OOP) lattice spacing is smaller than that of the substrate (higher $q_z$), consistent with the smaller unit cell volume for TmIG compared to GSGG. However, the thicker film (205 nm) is relaxed in plane with smaller IP and OOP lattice spacing than that of the substrate, and its peak position is close to that of bulk TmIG. The $\theta - 2\theta$ scans show a decrease in the OOP spacing (increase in $2\theta$) for thinner films. These trends are consistent with the TmIG initially growing with an IP lattice match to the substrate and hence a tensile IP strain (and a magnetoelastic anisotropy favoring PMA), but the strain relaxes as the film thickness increases. The thickest films, which are strain-relaxed, have a slightly higher OOP lattice spacing compared to bulk according to **Fig. 1**(a) which suggests the presence of oxygen vacancies or Tm:Fe ratio exceeding 0.6, which can occur in thin films and raise the unit cell volume. All the films show a smooth surface morphology with a low root-mean-square roughness below 0.5 nm, as visible in atomic force microscopy (AFM) images for the GSGG/TmIG(46nm), GGG/TmIG(44nm) and sGGG/TmIG(75nm) films shown in the **Supplementary Figure 1**.

A cross-section of an about 220 nm thick TmIG film on GSGG substrate, covered with a 5 nm Pt layer, was analyzed by scanning transmission electron microscopy (STEM). **Fig. 2**(a) shows a low magnification STEM image of the whole layer stack. An annular detector with a small collector angle (24-48 mrad) was used to highlight strain (Bragg) contrast over mass (Z) contrast [49]. The TmIG film shows columnar features attributed to strain contrast. An atomically resolved STEM image at the TmIG film-Pt interface (**Fig. 2**(b)) reveals a single crystalline TmIG film under the polycrystalline Pt layer, with the bright spots indicating columns of Tm and Fe. The STEM image of an area within the TmIG film close to the Pt interface shows the presence of a planar defect in which selected lattice planes are highlighted by colored lines in **Fig. 2**(c). Such planar defects could be associated with partial dislocations or atomic level disorder, which are common in REIGs.[50,51]

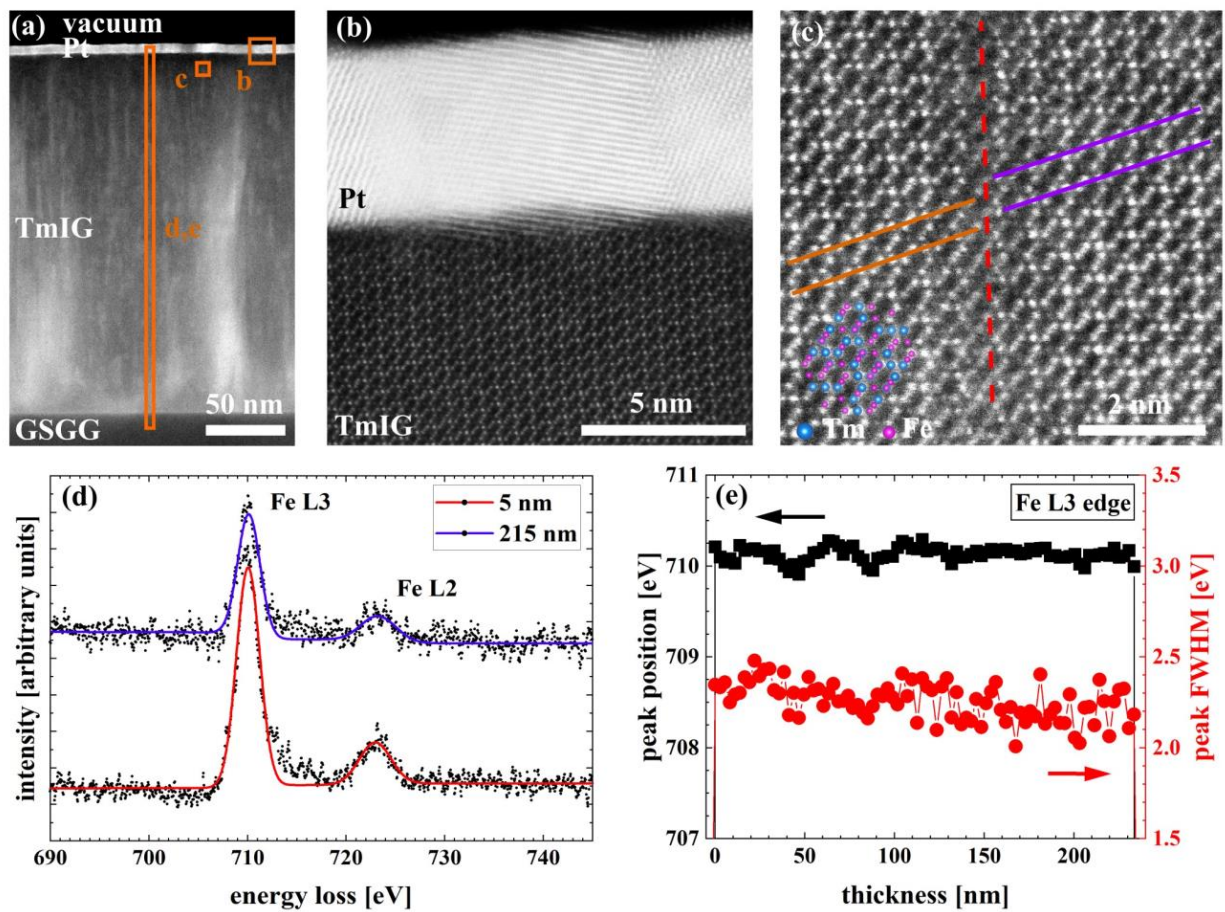

**Figure 2. Cross-sectional scanning transmission electron microscopy (STEM) analysis of the GSGG/TmIG(220 nm)/Pt(5nm) film.** (**a**) TEM image of the layer stack recorded by an annular detector with a small collector angle (24-48 mrad), highlighting strain (Bragg) contrast over mass (Z) contrast, (**b**) shows an atomic-resolution STEM image of the TmIG-Pt interface with [110] zone axis, while (**c**) shows an area within the TmIG film. The colored lines highlight a planar defect. (**d**) electron energy loss spectroscopy (EELS) scan at the Fe L3 and L2 edges. The measured energy loss spectra are displayed as data points, exemplified for positions close to the garnet-substrate and garnet-Pt interfaces, with the fitted functions presented as colored lines. (**e**) The thickness dependent Fe L3 peak position and FWHM is extracted.

## 2. 2. Correlation between Thermo-Spin Transport and Magnetism

**Fig. 3**(a) shows the schematic illustration of our LSSE measurement configuration. Simultaneous application of a vertical (+$z$-axis) $T$-gradient ($\overrightarrow{\nabla T}$) and an in-plane ($x$-axis) DC magnetic field ($\overrightarrow{\mu_0 H}$) across the TmIG film causes diffusion of thermally-excited magnons and develops a spatial gradient of magnon accumulation along the direction of $\overrightarrow{\nabla T}$.[52] The accumulated magnons close to the TmIG/Pt interface transfer spin angular momenta to the electrons of the adjacent Pt layer[52]. The injected spin current density is, $\overrightarrow{J_S} \propto -S_{LSSE}\overrightarrow{\nabla T}$, where $S_{LSSE}$ is the LSSE coefficient[52,53]. The spin current injected into the Pt layer along the $z$-direction is converted into a charge current, $\overrightarrow{J_C} = \left(\frac{2e}{\hbar}\right)\theta_{SH}^{Pt}(\overrightarrow{J_S} \times \overrightarrow{\sigma_S})$ along the $y$-direction via the inverse spin Hall effect (ISHE), where $e$, $\hbar$, $\theta_{SH}^{Pt}$, and $\overrightarrow{\sigma_S}$ are the electronic charge, the reduced Planck's constant, the spin Hall angle of Pt, and the spin-polarization vector, respectively. The corresponding LSSE voltage is[52,54,55]

$$V_{LSSE} = R_y L_y D_{Pt} \left(\frac{2e}{\hbar}\right)\theta_{SH}^{Pt}|J_S|\tanh\left(\frac{t_{Pt}}{2D_{Pt}}\right), \quad (1)$$

where, $R_y, L_y, D_{Pt},$ and $t_{Pt}$ represent the electrical resistance between the contact-leads, the distance between the contact-leads, the spin diffusion length of Pt, and the Pt layer thickness, respectively.

**Fig. 3**(b) shows the magnetic field ($H$) dependent ISHE voltage, $V_{ISHE}(H)$ for GSGG/TmIG(236 nm)/Pt(5 nm) for different values of the temperature difference between the hot ($T_{hot}$) and cold ($T_{cold}$) blocks, $\Delta T = (T_{hot} - T_{cold})$, at a fixed average sample temperature $T = \frac{T_{hot}+T_{cold}}{2} = 295$K. For all $\Delta T$, $V_{ISHE}(H)$ exhibits a nearly square-shaped hysteresis loop. The inset of **Fig. 3**(b) plots the $\Delta T$-dependence of the background-corrected LSSE voltage, $V_{LSSE}(\Delta T) = $

$\left[\frac{V_{ISHE}(+\mu_0 H_{sat}, \Delta T) - V_{ISHE}(-\mu_0 H_{sat}, \Delta T)}{2}\right]$, where $\mu_0 H_{sat}$ is the saturation field. Clearly, $V_{LSSE}$ increases linearly with $\Delta T$ as expected from **Eqn. 1**.[12]

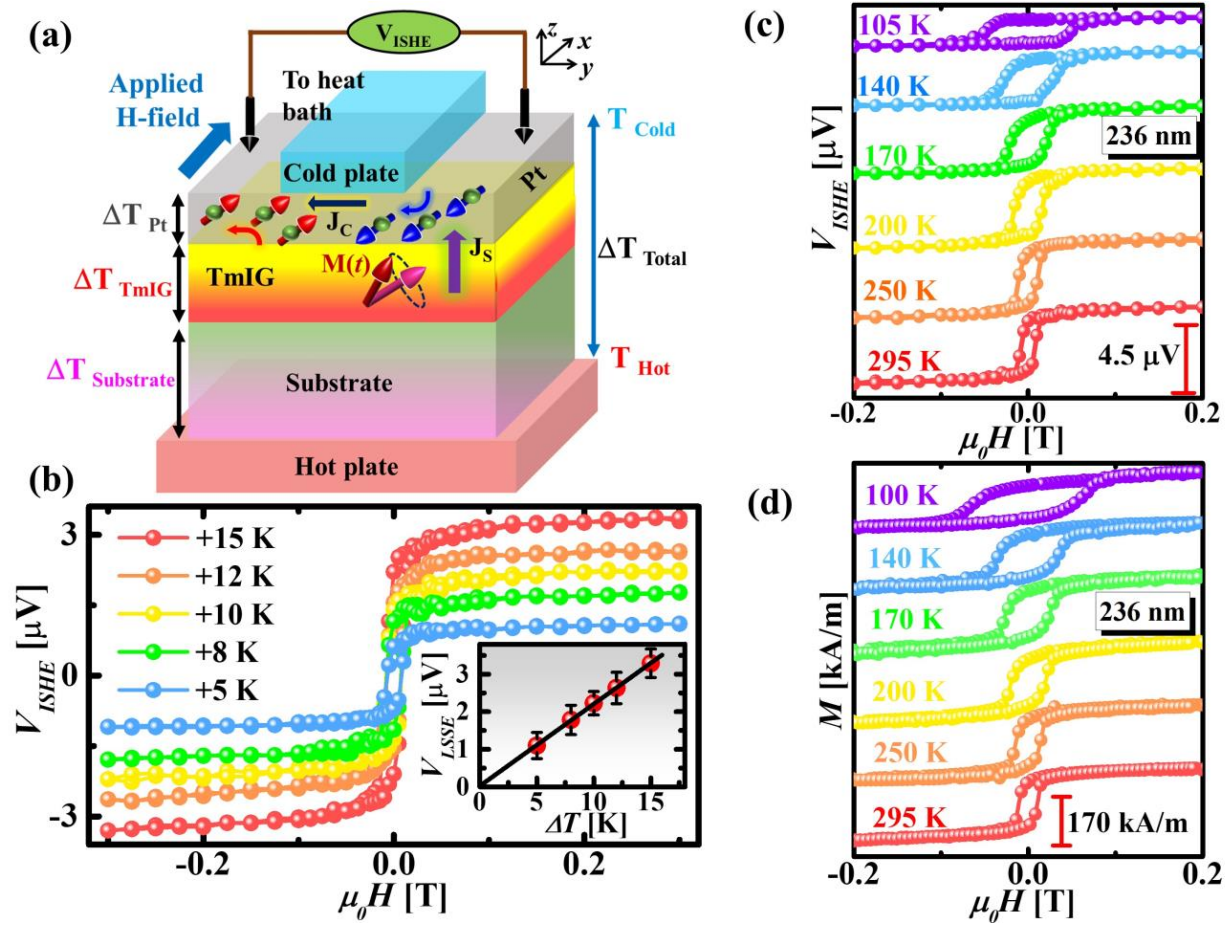

**Figure 3. Magnetism and longitudinal spin Seebeck effect (LSSE) in GSGG/TmIG(236nm)/Pt(5nm) film. (a)** Schematic illustration of the experimental configuration for LSSE measurements. A temperature gradient ($\vec{\nabla T}$) is applied along the +z axis and an in-plane (IP) dc magnetic field ($\vec{\mu_0 H}$) is applied along the +x axis. The inverse spin Hall effect (ISHE) induced voltage ($V_{ISHE}$) is measured along the y-axis. **(b)** $V_{ISHE}(H)$ loops for different values of the temperature difference $\Delta T$ at a fixed average sample temperature $T = 295$ K. The inset shows a linear $\Delta T$-dependence of the background-corrected LSSE voltage. **(c)** $V_{ISHE}(H)$ hysteresis loops measured at selected temperatures in the range 120 K $\leq T \leq$ 295 K for $\Delta T = +10$ K. **(d)** The IP $M(H)$ hysteresis loops at selected temperatures.

**Fig. 3**(c) shows the $V_{ISHE}(H)$ hysteresis loops for GSGG/TmIG(236 nm)/Pt(5 nm) measured at selected temperatures for $\Delta T$ = +10 K. Clearly, $|V_{ISHE}(\mu_0 H_{sat})|$ significantly decreases, and the hysteresis loop broadens at low temperatures, especially below 200 K. To correlate thermo-spin transport with the bulk magnetic properties, in **Fig. 3**(d), we show the magnetic field dependence of magnetization, $M(H)$ at selected temperatures for GSGG/TmIG(236 nm)/Pt(5 nm) measured while scanning an in-plane (IP) magnetic field. It is evident that with lowering the temperature, the saturation magnetization ($M_S$) decreases and the coercivity ($H_C$) increases with a corresponding increase in the magnetic anisotropy, especially below 200 K. This observation is also in agreement with the *T*-dependent magnetic force microscopy (MFM) results shown in **Supplementary Figure 2**, which clearly reveals that the root mean square (RMS) value of the phase shift, $\Delta\phi_{RMS}$ decreases significantly between 300 and 150 K indicating changes in the magnetic domain structure at low-*T*.

The decrease in $M_S$ at low-*T* is well-known in TmIG[48,56] and is a result of the increasing moment of the $Tm^{3+}$ ion at low-*T*, which competes with the net moment of the $Fe^{3+}$ ions (*i.e.*, the dodecahedral $Tm^{3+}$ moment opposes the net moment of the tetrahedral and octahedral $Fe^{3+}$ moments). Based on the molecular-field-coefficient theory developed by Dionne[57], we have performed molecular-field simulations[58,59] to determine $M_S(T)$ for TmIG (see **Supplementary Figure 3(n)**) which is consistent with our experimental observation of the decrease in $M_S$ at low-*T*. It is apparent from **Figs. 3**(c) and (d) that the temperature evolution of $V_{ISHE}$ signal follows that of $M_S$. To further explore the correlation between $V_{ISHE}$ and $M_S$, magnetometry and LSSE measurements were repeated on the GSGG/TmIG(*t*)/Pt(5 nm) sample series with different TmIG film thicknesses (28 nm $\leq t \leq$ 236 nm). Films with 46 nm $\leq t \leq$ 236 nm possess IP easy-axes while the 28 nm film has an OOP easy-axis of magnetization, which was confirmed via IP-magnetometry and OOP *p*-MOKE measurements (see **Supplementary Figure 3(e)**). The total magnetic anisotropy of a (111)-oriented TmIG film, neglecting growth and interfacial anisotropies, has contributions from shape anisotropy ($K_{shape}$), cubic magnetocrystalline anisotropy ($K_{mc}$), and magnetoelastic anisotropy ($K_{me}$)[47,49,60] *i.e.*, $K_{eff} = K_{shape} + K_{mc} + K_{me} = -\frac{1}{2}\mu_0 M_S^2 - \frac{K_1}{12} - \frac{9}{4}\lambda_{111}c_{44}\left(\frac{\pi}{2} - \beta\right)$, where $K_1$ is the magnetocrystalline anisotropy coefficient, $\lambda_{111}$ is the magnetostriction along the [111] direction, $c_{44}$ is the shear modulus and $\beta$ is the corner angle of the rhombohedrally-distorted unit cell. For a negative magnetostriction ($\lambda_{111} = -5.2 \times 10^{-6}$ for bulk TmIG[47]), the tensile IP strain, which results from the difference in lattice parameters ($a_{GSGG} = 12.57$ Å and $a_{TmIG} = 12.32$ Å) promotes PMA ($K_{eff} > 0$).[49,60,61] PMA is expected for fully-strained films (28 nm), but strain-relaxation in thicker films reduces the magnetoelastic contribution, and the easy-axis reorients to IP direction[60].

**Figs. 4**(a) and (b) depict the $V_{ISHE}(H)$ loop on the left *y*-scale and corresponding $M(H)$ loop on the right *y*-scale at 295 K for the thicknesses: $t = 236$ and 28 nm, respectively. The $M(H)$ and $V_{ISHE}(H)$ hysteresis-loops for all other thicknesses are shown in the **Supplementary Figures 3 and 4.** Clearly, the $V_{ISHE}(H)$ hysteresis-loops for all the thicknesses mimic the corresponding $M(H)$ loops. Note that, unlike YIG-slab, there is no surface magnetic anisotropy induced anomalous low field feature in the $V_{ISHE}(H)$ loop for any of our TmIG thin films. This is possibly because the thickness of the TmIG films is smaller than their average magnetic domain size[62]. This is why the YIG thin films also do not show any low field anomalous feature in the $V_{ISHE}(H)$ loops[52]. Additionally, the $V_{ISHE}(H)$ loop for our TmIG film with PMA (the 28 nm film) at 295 K is quite similar to that of a TmIG thin film with PMA at room temperature reported in the literature[46]. In **Figs. 4**(c) and (d), we demonstrate the $T$-dependence of the background-corrected LSSE voltage, $V_{LSSE}(T) = \frac{V_{ISHE}(T,+\mu_0 H_{sat}) - V_{ISHE}(T,-\mu_0 H_{sat})}{2}$ for $\Delta T = +10$ K on the left $y$-scale and corresponding $M_S(T)$ on the right $y$-scale for GSGG/TmIG(236 nm)/Pt(5 nm) and GSGG/TmIG(28 nm)/Pt(5 nm), respectively. Interestingly, $V_{LSSE}(T)$ and $M_S(T)$ for both films drop remarkably below the $T$-window of 180-200 K. We observed a similar trend in $V_{LSSE}(T)$ and $M_S(T)$ for all GSGG/TmIG($t$)/Pt(5 nm) films with other thicknesses (see **Supplementary Figures 3 and 4**). These results indicate that this behavior is intrinsic to TmIG.

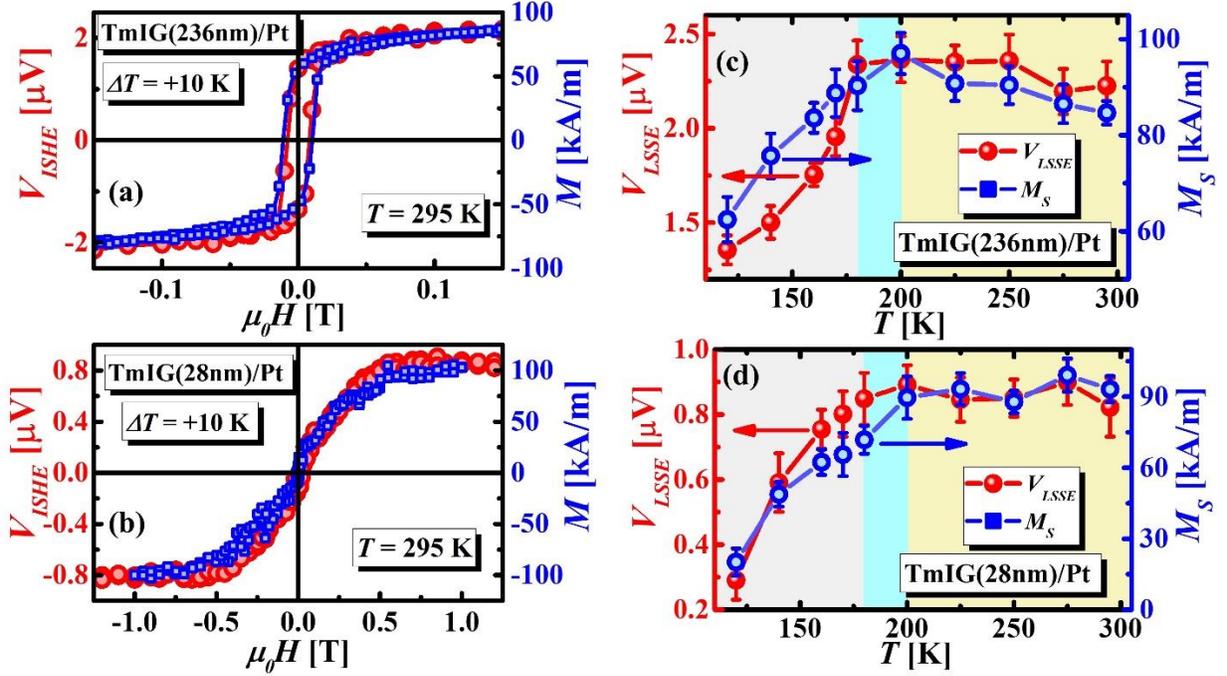

**Figure 4. Longitudinal spin Seebeck effect in GSGG/TmIG($t$)/Pt(5 nm) films.** The $V_{ISHE}(H)$ hysteresis loops on the left $y$-scale and the IP $M(H)$ loops on the right $y$-scale at $T = 295$ K for GSGG/TmIG($t$)/Pt films for $t = $ **(a)** 236 nm, and **(b)** 28 nm. The temperature dependence of the background-corrected LSSE voltage, $V_{LSSE}(T)$ on the left $y$-scale and temperature dependence of saturation magnetization, $M_S(T)$ on the right $y$-scale for the GSGG/TmIG($t$)/Pt(5 nm) films for $t = $ **(c)** 236 nm and **(d)** 28 nm, for $\Delta T = +10$ K.

Note that the yellow and grey background colors in all the graphs throughout the manuscript are used to highlight significant changes in physical parameters between high (yellow) and low (grey) temperature regions. Additionallly, we have used the sky blue background color in some of the specific graphs (especially temperature dependence of $V_{LSSE}(T)$ and $M_S(T)$) to indicate considerable changes in the corresponding physical parameters occurring around the narrow temperature window: 180 K $\leq T \leq$ 200 K. However, we have used a gradual transition from yellow to grey background in rest of the graphs where the changes in the physical parameters are less significant in the temperature window: 180 K $\leq T \leq$ 200 K.

Next, we discuss the additional voltage contributions due to the magnetic proximity effect (MPE)-induced anomalous Nernst effect (ANE) as well as MPE–induced LSSE in the Pt layer. The MPE leads to a magnetic moment in a few atomic layers of Pt close to the TmIG/Pt interface.[63,64] In the presence of a vertical temperature gradient, a transverse voltage is generated in the proximitized Pt layer due to ANE which adds to the LSSE voltage. Furthermore, due to the temperature gradient, spin currents are generated inside the magnetized Pt layer, which induces an additional IP charge current at the proximitized Pt/nonmagnetic Pt interface via the ISHE and therefore contributes to the LSSE signal.[65] In an earlier study, Bougiatioti *et al.*,[63] showed that the MPE-induced ANE in the proximitized Pt layer is only significant for a conducting FM/Pt bilayer but negligible for semiconducting FM/Pt bilayers and becomes zero for insulating FM/Pt bilayers. Since TmIG is insulating, the contribution of the MPE-induced ANE in the proximitized Pt layer towards the total LSSE signal can be neglected throughout the measured temperature range[66]. Furthermore, since the LSSE voltage decreases with decreasing thickness of the magnetic layer,[16] and the thickness of the proximitized Pt layer is very small, the MPE-induced LSSE contribution due to the proximitized Pt layer can also be neglected[66]. Therefore, the total voltage measured across the TmIG/Pt bilayers is considered to be solely contributed by the intrinsic LSSE of the TmIG films.

**2. 3. Analysis of the Thickness Dependent Longitudinal Spin Seebeck Effect**

To ascertain the origin of the decrease in $V_{LSSE}$ below 180-200 K in our TmIG films, it is essential to determine the temperature evolution of $\langle \xi \rangle$ which signifies the critical length-scale for the thermally-generated magnons of a magnetic thin film[16,18,34]. For an effective determination of the temperature dependence of $\langle \xi \rangle$, the contributions of the the thermal resistances of the substrate and the grease layers as well as the interfacial thermal resistances need to be considered.[67] To quantify the temperature evolution of $\langle \xi \rangle$ for our TmIG/Pt bilayer films, we have employed a model proposed by Jimenez-Cavero *et al*.,[68] according to which the total temperature difference ($\Delta T$) across the GSGG/TmIG/Pt heterostructure can be expressed as a linear combination of temperature drops in the Pt layer, at the TmIG/Pt interface, in the TmIG layer, at the GSGG/TmIG interface and across the GSGG substrate as well as in the N-grease layers (thickness ≈ 1 μm) on both sides of the GSGG/TmIG/Pt heterostructures as,[68] $\Delta T = \Delta T_{Pt} + \Delta T_{\frac{Pt}{TmIG}} + \Delta T_{TmIG} + \Delta T_{\frac{TmIG}{GSGG}} + \Delta T_{GSGG} + 2.\Delta T_{N-Grease}$ (see **Fig. 5**(a)). Assuming negligible drops in $\Delta T$ in the Pt layer and at the GSGG/TmIG and Pt/TmIG interface,[68,69] the total temperature difference can be approximately written as, $\Delta T = \Delta T_{\frac{Pt}{TmIG}} + \Delta T_{TmIG} + \Delta T_{GSGG} + 2.\Delta T_{N-Grease}$. Considering these contributions, the temperature drops in the TmIG layer and at the TmIG/Pt interface can be written as,[68,69] $\Delta T_{TmIG} = \frac{\Delta T}{\left[1+\frac{\kappa_{TmIG}}{t_{TmIG}}\left(\frac{2t_{N-Grease}}{\kappa_{N-Grease}}+\frac{t_{GSGG}}{\kappa_{GSGG}}\right)\right]}$ and $\Delta T_{\frac{Pt}{TmIG}} = \left[\frac{(\kappa_{GSGG}\kappa_{TmIG})R_{int}}{(\kappa_{TmIG}t_{GSGG}+\kappa_{GSGG}t_{TmIG})}\right]\Delta T$, respectively. The bulk $(V_{LSSE}^b)$ and interfacial $(V_{LSSE}^i)$ contributions to the LSSE voltage can then be expressed as, $V_{LSSE}^b = S_{LSSE}^b \cdot \Delta T_{\text{TmIG}} \cdot L_y = \left[\left(\frac{A}{t_{TmIG}}\right)\left\{\frac{\cosh\left(\frac{t_{TmIG}}{\langle\xi\rangle}\right)-1}{\sinh\left(\frac{t_{TmIG}}{\langle\xi\rangle}\right)}\right\}\right]\left\{\frac{\Delta T}{\left[1+\frac{\kappa_{TmIG}}{t_{TmIG}}\left(\frac{2t_{N-Grease}}{\kappa_{N-Grease}}+\frac{t_{GSGG}}{\kappa_{GSGG}}\right)\right]}\right\}L_y$ and $V_{LSSE}^i = S_{LSSE}^i \cdot \Delta T_{\frac{\text{Pt}}{\text{TmIG}}} \cdot L_y = S_{LSSE}^i \cdot \left[\frac{(\kappa_{GSGG}\kappa_{TmIG})R_{int}\Delta T}{(\kappa_{TmIG}t_{GSGG}+\kappa_{GSGG}t_{TmIG})}\right]L_y$, respectively. Here, $S_{LSSE}^b = \left[\left(\frac{A}{t_{TmIG}}\right)\left\{\frac{\cosh\left(\frac{t_{TmIG}}{\langle\xi\rangle}\right)-1}{\sinh\left(\frac{t_{TmIG}}{\langle\xi\rangle}\right)}\right\}\right]$ and $S_{int}$ denote the bulk and interfacial LSSE coefficients for TmIG and TmIG/Pt interface, respectively, $t_{TmIG}$ ($t_{GSGG}$) is the thickness of TmIG film (GSGG substrate), $\kappa_{TmIG}$ and $\kappa_{GSGG}$ are the thermal conductivity of TmIG and GSGG respectively, $t_{N-Grease}$ and $\kappa_{N-Grease}$ are the thickness and thermal conductivity of the N-grease layers, $R_{int}$ is the interfacial thermal-resistance at the TmIG/Pt interface and $A$ is a constant.[68]. The approximate values of $\kappa_{N-Grease}$, $\kappa_{TmIG}$ and $\kappa_{GSGG}$ at different temperatures are obtained from the literature[70–75]. Note that, we have ignored the interfacial thermal resistances between the N-grease and the hot/cold plates as well as between the sample and N-grease layers.[76] Therefore, the total LSSE voltage across GSGG/TmIG/Pt can be expressed as,[68]

$$V_{LSSE}(t_{TmIG}) = V_{LSSE}^i(t_{TmIG}) + V_{LSSE}^b(t_{TmIG}) = \left[S_{int}\left\{\frac{(\kappa_{GSGG}\kappa_{TmIG})R_{int}}{(\kappa_{TmIG}t_{GSGG}+\kappa_{GSGG}t_{TmIG})}\right\}L_y\Delta T + \left[\left(\frac{A}{t_{TmIG}}\right)\left\{\frac{\cosh\left(\frac{t_{TmIG}}{\langle\xi\rangle}\right)-1}{\sinh\left(\frac{t_{TmIG}}{\langle\xi\rangle}\right)}\right\}\right]\left\{\frac{\Delta T}{\left[1+\frac{\kappa_{TmIG}}{t_{TmIG}}\left(\frac{2t_{N-Grease}}{\kappa_{N-Grease}}+\frac{t_{GSGG}}{\kappa_{GSGG}}\right)\right]}\right\}L_y\right] \quad (2)$$

In **Fig. 5**(b), we demonstrate the $V_{ISHE}(H)$ hysteresis loops at $T = 295$ K for the GSGG/TmIG/Pt films with different $t_{TmIG}$ in the range $28\,\text{nm} \leq t \leq 236\,\text{nm}$. Clearly, $|V_{ISHE}(\mu_0 H_{sat})|$ decreases significantly with decreasing $t_{TmIG}$. Therefore, we fitted the thickness dependent LSSE voltage at different temperatures with **Eqn. 2** to evaluate the temperature dependence of $\langle\xi\rangle$ for our GSGG/TmIG/Pt films. It has recently been shown[77] that the $M_S$ also needs be considered to evaluate $\langle \xi \rangle$ from the LSSE voltage by normalizing the LSSE voltage by $M_S$. In **Fig. 5**(c), we show the thickness-dependence of the background-corrected modified LSSE voltage, $\frac{V_{LSSE}(t_{TmIG})}{\Delta T . M_S}$, at selected temperatures fitted to **Eqn. 2.** From the fits, we obtained $\langle \xi \rangle = 62 \pm 5$ nm for the TmIG film at 295 K, which is smaller than that of YIG thin films grown by PLD (90–140 nm)[16], but higher than that for GdIG thin films (45±8 nm)[12].

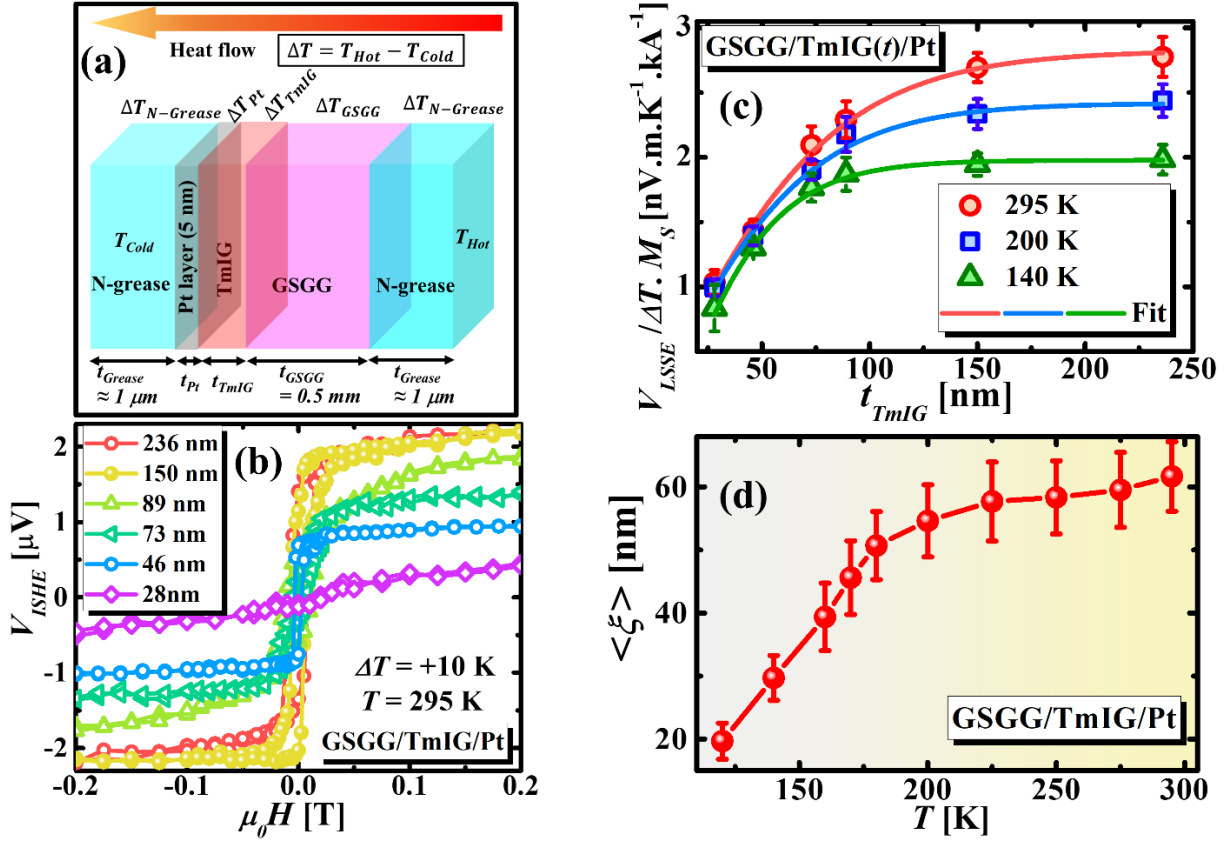

**Figure 5. Thickness Dependent LSSE and Magnon Propagation Length in GSGG/TmIG(*t*)/Pt(5 nm) films.** **(a)** Schematic illustration of heat flow across the GSGG/TmIG(*t*)/Pt(5 nm) films. **(b)** The $V_{ISHE}(H)$ hysteresis loops for GSGG/TmIG(*t*)/Pt films with different thicknesses at $T = 295$ K for $\Delta T = +10$ K. **(c)** The thickness dependence of the normalized background corrected LSSE voltage, $V_{LSSE}(t)/\Delta T . M_S$ at three selected temperatures $T = 295, 200,$ and 140 K fitted with Eqn. (2). **(d)** The temperature dependence of the magnon propagation length, $\langle \xi \rangle$ obtained from the fits.

**Fig. 5**(d) demonstrates the *T*-dependence of $\langle\xi\rangle$ obtained from the fit of $\frac{V_{LSSE}(t_{TmIG})}{\Delta T.M_S}$ for GSGG/TmIG(*t*)/Pt(5nm) films. Interestingly, $\langle\xi\rangle$ decreases gradually with decreasing temperature and shows a comparatively faster decrease at low temperatures, especially below 200 K. Our observation is strikingly different than that reported by Guo *et al.*[34] for YIG/Pt bilayers. From the YIG thickness dependence of the local LSSE measurements in YIG/Pt, they determined the temperature dependence of $\langle\xi\rangle$ and found a scaling behavior of $\langle\xi\rangle \propto T^{-1}$.[34] However, by employing nonlocal measurement geometries, Cornelissen *et al.*, demonstrated that $\langle\xi\rangle$ (and hence, the magnon diffusion length) for YIG/Pt decreases with decreasing temperature over a broad temperature range[35], similar to what we observed in our TmIG/Pt bilayers. Gomez-Perez *et al.*,[19] also observed similar behavior of the magnon diffusion length in YIG/Pt. However, none of these studies indicated significant change in $\langle\xi\rangle$ at low temperatures. Therefore, the observed temperature evolution of $\langle\xi\rangle$ presented in this study is intrinsic to TmIG. To rule out possible effects of strain on $V_{LSSE}(T)$, we performed LSSE measurements on TmIG films grown on different substrates (see **Supplementary Figures 5 and 6**). It is evident that $M_S(T)$ and $V_{LSSE}(T)$ for the $Gd_3Ga_5O_{12}$(GGG)/TmIG(44 nm)/Pt(5 nm) and $(Gd_{2.6}Ca_{0.4})(Ga_{4.1}Mg_{0.25}Zr_{0.65})O_{12}$(sGGG)/TmIG(40 nm)/Pt(5 nm) films (see **Supplementary Figure 7**) exhibit the same trend as GSGG/TmIG(46 nm)/Pt(5 nm). More specifically, both $V_{LSSE}(T)$ and $M_S(T)$ drop below 180-200 K for all the TmIG films independent of substrate choice.

To interpret the decrease in $\langle\xi\rangle$ at low temperatures, we recall that $\langle\xi\rangle$ of a magnetic material with lattice constant $a_0$ (considering simple cubic structure) is related to the Gilbert damping parameter ($\alpha$), the effective magnetic anisotropy constant ($K_{eff}$), and the strength of the Heisenberg exchange interaction between nearest neighbors ($J_{ex}$) through the relation[17,18] $\langle \xi \rangle = \frac{a_0}{2\alpha} \cdot \sqrt{\frac{J_{ex}}{2K_{eff}}}$. As discussed before, $K_{eff} = K_{me} - \frac{1}{2}\mu_0 M_S^2 - \frac{K_1}{12}$. Therefore, we can express $\langle \xi \rangle$ as,

$$\langle \xi \rangle = \frac{a_0}{2\alpha} \cdot \sqrt{\frac{J_{ex}}{2\left(K_{me} - \frac{K_1}{12} - \frac{1}{2}\mu_0 M_S^2\right)}} \tag{3}$$

**Eqn. 3** indicates that (i) $\langle \xi \rangle \propto \left(\frac{1}{\alpha}\right)$, and (ii) a decrease in $M_S$ also suppresses $\langle \xi \rangle$. Since the effective anisotropy field, $H_K^{eff} \propto K_{eff}$, **Eqn. 3** can be alternatively written as $\langle \xi \rangle = \frac{a_0}{2\alpha} \cdot \sqrt{\frac{J_{ex}}{2K_{eff}}} \propto \frac{1}{\alpha \cdot \left(H_K^{eff}\right)^{1/2}}$, which indicates that $\langle \xi \rangle$ is inversely proportional to the square-root of $H_K^{eff}$. This implies that the temperature evolution of $\langle \xi \rangle$ is intrinsically dependent on both the physical quantities: $\alpha$ and $H_K^{eff}$. To determine the roles of $\alpha$ and $H_K^{eff}$ in the temperature evolution of $\langle \xi \rangle$, we have performed radio frequency (RF) transverse susceptibility (TS) and broadband ferromagnetic resonance (FMR) measurements, respectively on the TmIG films, which have been discussed in the following sections.

## 2. 4. Radio Frequency Transverse Susceptibility and Magnetic Anisotropy

RF TS measurements were performed to determine the temperature evolution of $H_K^{eff}$ in the TmIG films. The magnetic field dependence ($H_{DC}$) of TS, $\chi_T(H_{DC})$, is known to exhibit peaks/cusps at the effective anisotropy fields, $\pm H_K^{eff}$.[78,79] The schematic illustration of our TS measurement configuration is shown in **Fig. 6**(a). The RF magnetic field, $H_{RF}$ is parallel to the film surface and $H_{DC}$ points perpendicular to it. All the TS data in this paper are presented as the relative change in $\chi_T(H_{DC})$, which we define as $\frac{\Delta \chi_T}{\chi_T}(H_{DC}) = \frac{\chi_T(H_{DC}) - \chi_T(H_{DC} = H_{DC}^{sat})}{\chi_T(H_{DC}^{sat})}$, where $\chi_T(H_{DC} = H_{DC}^{sat})$ is the value of $\chi_T(H_{DC})$ at the saturation field ($H_{DC}^{sat}$). Bipolar field-scans of $\frac{\Delta \chi_T}{\chi_T}(H_{DC})$ for the GSGG/TmIG(236 nm)/Pt film at 295 and 100 K are shown in **Fig. 6**(b), which clearly indicates an increase in $H_K^{eff}$ at low-$T$.

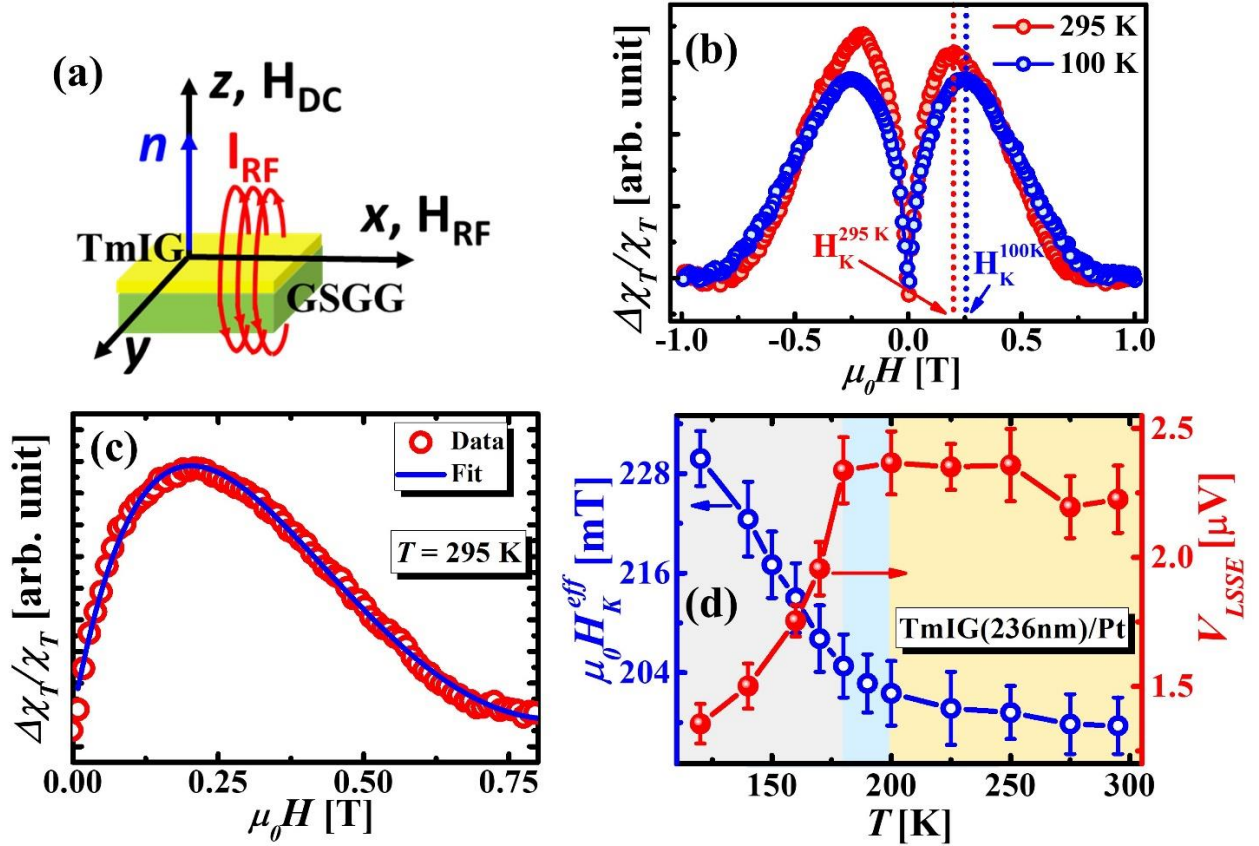

**Figure 6. RF Transverse Susceptibility and Magnetic Anisotropy in GSGG/TmIG(*t*)/Pt(5 nm) films.** **(a)** The schematic illustration of our RF transverse susceptibility measurement. **(b)** Comparison of the bipolar field scans ($+H_{DC}^{max} \rightarrow -H_{DC}^{max} \rightarrow +H_{DC}^{max}$) of transverse susceptibility at $T$ = 295 and 100 K for the GSGG/TmIG(236 nm)/Pt film measured with configuration $H_{DC} \perp$ film surface (IP easy axis). **(c)** Fitting of our $\frac{\Delta \chi_T}{\chi_T}(H_{DC})$ data for the GSGG/TmIG(236nm)/Pt film at 295 K with the Eqn. 4. **(d)** Temperature dependence of the effective anisotropy field ($H_K^{eff}$) for the GSGG/TmIG(236 nm)/Pt(5 nm) film obtained from the transverse susceptibility (TS) measurements on the left *y*-scale and corresponding $V_{LSSE}(T)$ for the same film on the right *y*-scale.

For an accurate determination of $H_K^{eff}$ from the field dependent TS curves, we fitted the line shapes for the TS curves with the following expression,[79,80]

$$\frac{\Delta \chi_T}{\chi_T}(H_{DC}) = \Delta \chi_{Sym} \frac{\left(\frac{\Delta H}{2}\right)^2}{\left(H_{DC}-H_K^{eff}\right)^2 + \left(\frac{\Delta H}{2}\right)^2} + \Delta \chi_{Asym} \frac{\frac{\Delta H}{2}\left(H_{DC}-H_K^{eff}\right)}{\left(H_{DC}-H_K^{eff}\right)^2 + \left(\frac{\Delta H}{2}\right)^2} + \Delta \chi_0 \qquad (4)$$

where, $\Delta H$ is the linewidth of the $\frac{\Delta \chi_T}{\chi_T}(H_{DC})$ spectrum, $\Delta \chi_{Sym}$ and $\Delta \chi_{Asym}$ are the coefficients of symmetric and antisymmetric Lorentzian functions and $\Delta \chi_0$ is the constant offset parameter. **Fig. 6**(c) shows the fitting of our $\frac{\Delta \chi_T}{\chi_T}(H_{DC})$ data for the GSGG/TmIG(236 nm)/Pt film at 295 K with the **Eqn. 4**. As shown on the left *y*-scale of **Fig. 6**(d), $H_K^{eff}(T)$ increases throughout the measured temperature range but the increase in $H_K^{eff}$ is comparatively faster below the temperature range: 180-200 K, which coincides with the remarkable drop in $V_{LSSE}$. Similar behavior was also observed for other film thicknesses (see **Supplementary Figure 8**).

A significant increase in magnetocrystalline anisotropy at low-*T* has been reported in various REIGs, which was interpreted in the framework of the single-ion anisotropy model considering the collective influence of the crystal and exchange fields of the REIG on the energy levels of the individual magnetic ions[81]. Typically, $K_1$ increases by ≈ 80-100% between 300 and 150 K in most of the REIGs.[81] Furthermore, $\lambda_{111}$ for TmIG increases from $-5.2 \times 10^{-6}$ at 300 K to $-17.4 \times 10^{-6}$ at 150 K which gives rise to enhanced contribution of $K_{me}$ towards $K_{eff}$ in TmIG films at low temperatures.[82,83] Shumate Jr. *et al.*,[84] observed a rapid increase in $H_K^{eff}$ and coercive field at low temperatures in mixed REIGs. An increase in $H_K^{eff}$ and a corresponding decrease in $V_{LSSE}$ below 175K was also observed in YIG/Pt[13], which was attributed to the single-ion anisotropy of $Fe^{2+}$ ions[85]. To gain knowledge on the oxidation state of Fe in our TmIG films, electron energy loss spectroscopy (EELS) was conducted during the cross-sectional TEM study described earlier. **Fig. 2**(d) shows two EELS spectra, recorded at the Fe L3 and L2 edges, and at positions close to the film-substrate and the film-Pt interface. The spectra are fitted following[86,87], shown by colored lines, using a Gaußian profile and a combination of a power-law background and a double-step function (arctangent) with a fixed step-ratio. **Fig. 2**(e) shows the extracted thickness-dependent Fe L3 peak position alongside the corresponding FWHM. While an exact quantification of the Fe oxidation state distribution using the EELS Fe L3 peak position or L3/L2 white-line ratio is challenging, the presence of different oxidation states can be indicated qualitatively by a shift in the peak position because $Fe^{2+}$ ions contribute at slightly lower energies compared to $Fe^{3+}$ ions[86–89]. However, in our measured spectra, a constant peak position at about 710.1 eV and a constant FWHM of about 2.3 eV across the whole film thickness is observed. Our observation strongly hints at the presence of only one Fe oxidation state, namely the $Fe^{3+}$ ion and hence, we can rule out the contribution of single ion anisotropy of $Fe^{2+}$ ions towards the increased magnetic anisotropy. This is also in agreement with recent studies on Tb-rich TbIG thin films[58,90] which reveal very low $Fe^{2+}$ ion concentrations. Therefore, the increase in $H_K^{eff}$ below 200 K in the TmIG films may arise from single-ion anisotropies of the $Tm^{3+}$ and $Fe^{3+}$ ions[81] as well as from the enhanced contributions of $K_1$ and $K_{me}$ towards $K_{eff}$ at low temperatures[81–83].

## 2. 5. Magnetization Dynamics and Broadband Ferromagnetic Resonance

Next, we examine the temperature evolution of $\alpha$ and its influence on $\langle\xi\rangle$ through broadband IP FMR measurements. **Fig. 7**(a) shows the field-derivative of the microwave (MW) power absorption spectra $\left(\frac{dP}{dH}\right)$ as a function of the IP DC magnetic field for a fixed frequency $f = $ 12 GHz at selected temperatures for the GSGG/TmIG(236nm) film. As temperature decreases, the $\left(\frac{dP}{dH}\right)$ lineshape noticeably broadens and the resonance field $H_{res}$ shifts to higher field values. The linewidth of the $\left(\frac{dP}{dH}\right)$ lineshape becomes so broad at low temperatures that we were unable to detect the FMR signal below 160 K. We observed the same behavior for the GSGG/TmIG(236 nm)/Pt(5 nm) film, as shown in the **Supplementary Figure 9**. **Fig. 7**(b) shows the $\left(\frac{dP}{dH}\right)$ lineshapes for the GSGG/TmIG(236 nm) film for different frequencies in the range 6 GHz $\leq f \leq$ 20 GHz at 295K fitted with a linear combination of symmetric and antisymmetric Lorentzian function derivatives as,[91]

$$\frac{dP}{dH} = P_{Sym} \frac{\frac{\Delta H}{2}(H_{dc}-H_{res})}{\left[(H_{dc}-H_{res})^2+\left(\frac{\Delta H}{2}\right)^2\right]^2} + P_{Asym} \frac{\left(\frac{\Delta H}{2}\right)^2-(H_{dc}-H_{res})^2}{\left[(H_{dc}-H_{res})^2+\left(\frac{\Delta H}{2}\right)^2\right]^2} + P_0 \qquad (5)$$

where, $H_{res}$ is the resonance field, $\Delta H$ is the linewidth of the $\frac{dP}{dH}$ lineshapes, $P_{Sym}$ and $P_{Asym}$ are the coefficients of the symmetric and antisymmetric Lorentzian derivatives, respectively, and $P_0$ is a constant offset parameter. The fitted curves are shown by solid lines in **Fig. 7**(b). Using the values of $H_{res}$ obtained from the fitting of the $\frac{dP}{dH}$ lineshapes, we fitted the $f$-$H_{res}$ curves at different temperatures using the Kittel equation for magnetic thin films with IP magnetic field,[92] which is expressed as $f = \frac{\gamma\mu_0}{2\pi}\sqrt{H_{res}(H_{res}+M_{eff})}$, where $M_{eff}$ is the effective magnetization, $\frac{\gamma}{2\pi} = \frac{g_{eff}\mu_B}{\hbar}$ is the gyromagnetic ratio, $\mu_B$ is the Bohr magneton, $g_{eff}$ is the effective Landé $g$-factor, and $\hbar$ is the reduced Planck's constant. **Fig. 7**(d) demonstrates the fitting of the $f$-$H_{res}$ curves at $T$ = 295, 200, and 160 K. We found that $g_{eff}$ = 1.642 ± 0.002 at $T$ = 295 K for our GSGG/TmIG(236 nm) film, which is significantly lower than that of the free electron value ($g_{eff}$ = 2.002), but close to the bulk TmIG value ($g_{eff}$ = 1.63)[93] as well as that for TmIG thin films ($g_{eff} \approx$ 1.57)[6,94]. Furthermore, as shown in **Fig. 7**(e), $g_{eff}$ for our GSGG/TmIG(236 nm) film decreases gradually with decreasing temperature. We observed similar behavior of $g_{eff}$ for the GSGG/TmIG(236 nm)/Pt(5 nm) film, and well as for other TmIG film thicknesses (see **Supplementary Figures 9 and 10**).

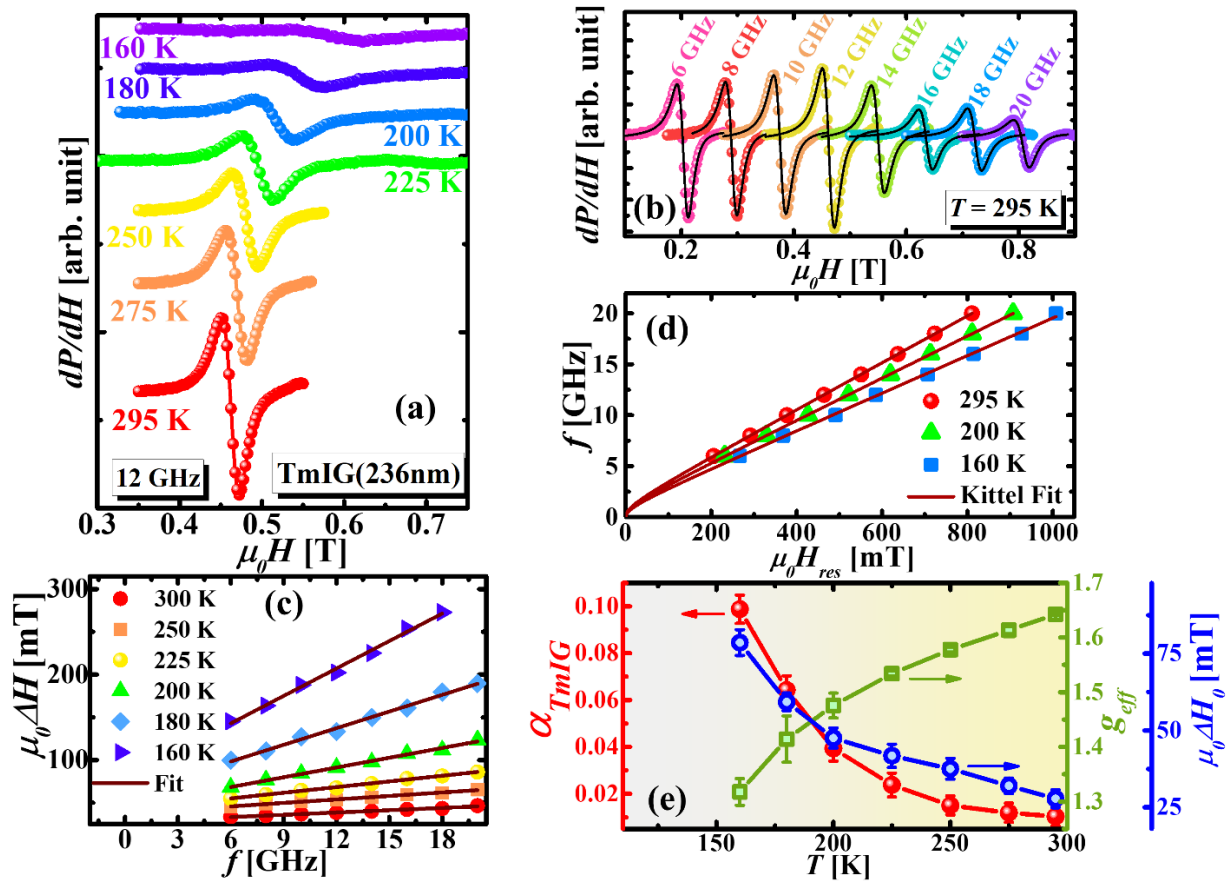

**Figure 7. Broadband Ferromagnetic Resonance.** (**a**) The field derivative of microwave (MW) power absorption spectra $\left(\frac{dP}{dH} \text{ line shapes}\right)$ for the GSGG/TmIG(236 nm) film at a fixed frequency ($f$ = 12 GHz) in the range 160 K ≤ $T$ ≤ 295 K. (**b**) $\frac{dP}{dH}$ line shapes at different frequencies between $f$ = 6 - 20 GHz fitted with the linear combination of symmetric and anti-symmetric Lorentzian function derivatives for the GSGG/TmIG(236 nm) film at $T$ = 295 K. (**c**) Frequency dependence of linewidth, $\Delta H$ at different temperatures for the GSGG/TmIG(236 nm) film with linear fit. (**d**) The $f$-$H_{res}$ curves at $T$ = 295, 200, and 160 K along with Kittel fits. (**e**) Temperature dependence of the Gilbert damping parameter, $\alpha_{TmIG}$, the inhomogeneous broadening, $\Delta H_0$ and the effective Landé $g$-factor for the GSGG/TmIG(236 nm) film.

Finally, to quantify the temperature dependence of the Gilbert damping parameter ($\alpha_{TmIG}$), we fitted the $\Delta H$-$f$ curves at different temperatures using the expression,[95] $\Delta H = \Delta H_0 + \frac{4\pi\alpha}{\gamma\mu_0}f$, where $\Delta H_0$ is the frequency-independent contribution to the linewidth, known as the inhomogeneous broadening linewidth. From the fits (see **Fig. 7**(c)), we obtained $\alpha_{TmIG}= 0.0103 \pm 0.002$ at 295 K for our GSGG/TmIG(236 nm) film which is close to the previously reported values of $\alpha$ ($\approx$ 0.0132-0.0146) for TmIG films[6,96]. Most importantly, $\alpha_{TmIG}$ increases gradually with decreasing temperature but shows a comparatively faster increase at low temperatures, especially below $\approx$ 200 K (**Fig. 7**(e)). A similar increase in $\alpha$ at low-$T$ has also been observed in GSGG/TmIG(236 nm)/Pt(5 nm), GSGG/TmIG(46 nm)/Pt(5 nm) and GGG/TmIG(44 nm)/Pt(5 nm) films (see **Supplementary Figures 9 and 10**), indicating that this behavior is independent of TmIG film thickness and substrate choice. In compensated ferrimagnetic insulators, *e.g.*, GdIG, $\alpha$ increases drastically close to the magnetic compensation temperature[32]. However, most of the earlier reports indicate that TmIG films do not show magnetic compensation in the temperature range between 1.5 and 300 K.[47,48] Since our TmIG films also do not show magnetic compensation in the measured temperature range, the increased value of $\alpha_{TmIG}$ at low temperatures in our TmIG films has a different origin. Sizeable increases in $\alpha$ and $\Delta H$ at low temperatures were also observed in YIG and different REIGs[92,97–99] including TmIG[46], which was primarily attributed to $Fe^{2+}$ and/or $RE^{3+}$ impurity relaxation mechanisms. However, our EELS study confirms the absence of $Fe^{2+}$ ions, and therefore, we can rule out the possibility of $Fe^{2+}$ impurity relaxation in our TmIG films. Therefore, the increased damping at low temperatures in our TmIG films may be associated with enhanced magnon scattering by defects,[100–104] and slowly relaxing $Tm^{3+}$ ions[92,105]. It is known that the contribution of slowly relaxing RE impurity ions towards damping is proportional to the orbital moment ($L$) of the $RE^{3+}$ ions[105–107], suggesting that this mechanism applies to $Tm^{3+}$ ($L = 5$).

## 2. 6. Correlating Magnon Propagation Length with Magnetic Anisotropy and Gilbert Damping

In the previous sections, we have demonstrated that both $H_K^{eff}$ and $\alpha_{TmIG}$ for our TmIG films show clear increases at low temperatures, especially below 200 K. It is known that the magnon energy-gap ($\hbar\omega_M$) is related to $K_{eff}$ through the expression: $\hbar\omega_M \propto 2K_{eff}$ [17,18]. Therefore, an increase in $H_K^{eff}$ (and hence, $K_{eff}$) below 200 K enhances $\hbar\omega_M$ giving rise to only high-frequency magnon propagation with shorter $\langle\xi\rangle$. Since only the subthermal magnons, *i.e.*, the low frequency magnons are primarily responsible for the long-range thermo-spin transport and contributes towards LSSE[19–21], the $V_{LSSE}$ signal also decreases below 200 K in our TmIG films[12,13]. This also explains the noticeable decrease in $\langle\xi\rangle$ below 200 K, as the maximum value of the frequency-dependent propagation length is $\langle\xi\rangle_{max} \propto \frac{1}{\sqrt{\hbar\omega_M^{min}}}$, where $\hbar\omega_M^{min}$ is the minimum value of $\hbar\omega_M$, and $\hbar\omega_M^{min} \propto 2K_{eff}$.[17] Therefore, according to the expression[17,18] $\langle\xi\rangle = \frac{a_0}{2\alpha} \cdot \sqrt{\frac{J_{ex}}{2K_{eff}}} \propto \frac{1}{\alpha \cdot (H_K^{eff})^{1/2}}$, the observed decrease in $\langle\xi\rangle$ and hence, the $V_{LSSE}$ signal at low temperatures, especially below 200 K in our TmIG films has contributions from the temperature evolutions of both $H_K^{eff}$ and $\alpha_{TmIG}$. The roles of magnetic anisotropy and damping in LSSE in different REIG-based MI/HM bilayers have been explored by different groups[12,13,32,33]. All these studies indicated that the LSSE signal strength varies inversely with both magnetic anisotropy and damping. In this manuscript, we have not only highlighted the roles of the temperature evolutions of both magnetic anisotropy and damping in controlling the temperature dependent LSSE effect in TmIG/Pt bilayers, but also attempted to establish possible correlations between $\langle\xi\rangle$, $H_K^{eff}$ and $\alpha$. Since $\langle\xi\rangle$ is intrinsic to a magnetic film and hence independent of the thickness of the magnetic film[19], it is convenient to directly correlate $\langle \xi \rangle$ with the physical parameters $H_K^{eff}$ and $\alpha$ of individual magnetic films with different thicknesses. We display the temperature dependence of $\langle \xi \rangle$ on the left-*y* scales, and the temperature evolutions of $H_K^{eff}$ and $\alpha_{TmIG}$ are shown on the right *y*-scales of **Figs. 8**(a) and (b), respectively. It is evident that the prominent drop in $\langle \xi \rangle$ below 200 K in the TmIG/Pt bilayers is associated with the noticeable increases in $H_K^{eff}$ and $\alpha$ that occur within the same temperature range.

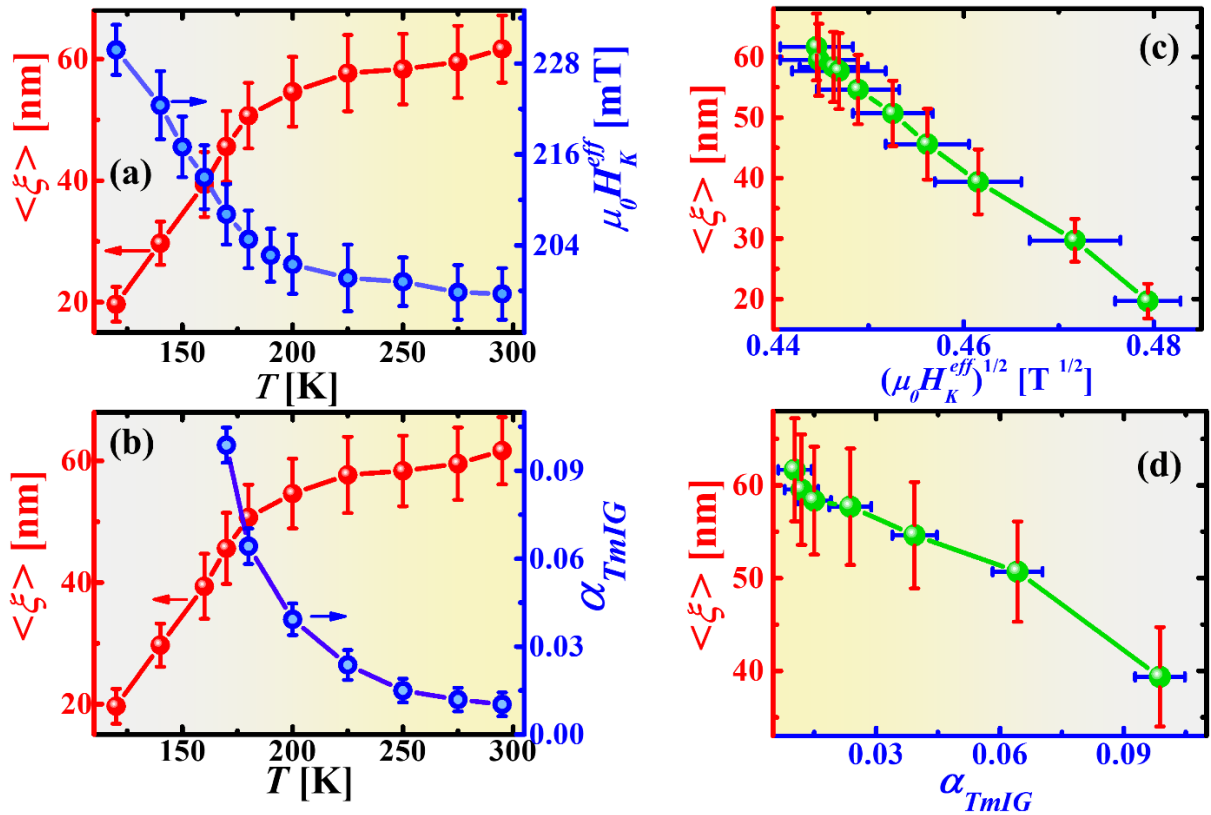

**Figure 8. Temperature evolution of magnon propagation length and its correlation with magnetic anisotropy and Gilbert damping: (a) and (b)** Temperature dependence of $\langle \xi \rangle$ on the left-*y* scales, and the temperature evolutions of $H_K^{eff}$ and $\alpha_{TmIG}$ are shown on the right *y*-scales, respectively. $\langle \xi \rangle$ as a function of **(c)** $\sqrt{H_K^{eff}}$ and **(d)** $\alpha_{TmIG}$ for the GSGG/TmIG(236 nm) film obtained from the temperature evolutions of $\langle \xi \rangle$, $H_K^{eff}$ and $\alpha_{TmIG}$.

For a clearer understanding of the direct correlation between $\langle\xi\rangle$ and $H_K^{eff}$ in our TmIG/Pt bilayer films, we have plotted $\langle\xi\rangle$ as a function of $\sqrt{H_K^{eff}}$ for the GSGG/TmIG(236 nm)/Pt film in **Fig. 8**(c) obtained from the temperature evolutions of $\langle\xi\rangle$ and $H_K^{eff}$. $\langle\xi\rangle$ varies inversely with $\sqrt{H_K^{eff}}$ in the measured temperature range, which is consistent with the expression $\langle\xi\rangle \propto \frac{1}{\alpha\cdot\left(H_K^{eff}\right)^{1/2}}$. Similarly, we have plotted $\langle\xi\rangle$ as a function of $\alpha_{TmIG}$ for the GSGG/TmIG(236 nm)/Pt film in **Fig. 8**(d) obtained from the temperature evolutions of $\langle\xi\rangle$ and $\alpha_{TmIG}$. An inverse correlation between $\langle\xi\rangle$ and $\alpha_{TmIG}$ in the measured temperature range is also apparent from this plot, and hence, in agreement with the aforementioned theoretical expression. To establish a more accurate correlation between the parameters $\langle\xi\rangle$, $H_K^{eff}$ and $\alpha$, one needs to fix $H_K^{eff}(\alpha)$, and then evaluate $\langle\xi\rangle$ for different values of $\alpha\left(H_K^{eff}\right)$. It is however challenging to change $H_K^{eff}$ of a magnetic material without varying $\alpha$ significantly. Nevertheless, we have observed concurrent remarkable drops in the LSSE voltage as well as $\langle\xi\rangle$ below 200 K in our TmIG/Pt bilayers regardless of TmIG film thickness and substrate choice and correlated the temperature evolution of $\langle\xi\rangle$ with the noticeable increases in $H_K^{eff}$ and $\alpha$ that occur within the same temperature range. It is important to note that FMR probes only the zone center magnons in the GHz range whereas the subthermal magnons which primarily contribute towards the LSSE signal belong to the THz regime[34]. Therefore, the behavior of LSSE cannot be determined by FMR excited by GHz-range microwaves. As shown by Chang *et al.*,[33] the LSSE voltage varies inversely with $\alpha$. Furthermore, the correlation between $\langle\xi\rangle$ and $\alpha$ was predicted theoretically[18] but never shown experimentally. As indicated in this report, an experimental demonstration of the correlation between $\langle\xi\rangle$, $H_K^{eff}$ and $\alpha$ would be beneficial to fabricate efficient spincaloritronic devices with higher $\langle\xi\rangle$ by tuning these fundamental parameters. However, for deeper understanding of LSSE and $\langle\xi\rangle$, their magnon frequency dependences need to be highlighted.

**2. 7. Magnon Frequency Dependences of the LSSE Voltage and Magnon Propagation Length**

As discussed before, the low energy subthermal magnons with longer $\langle\xi\rangle$ are primarily responsible for LSSE. These low frequency magnons are partially frozen out by the application of external magnetic field because of increased magnon energy gap due to the Zeeman effect.[20,108] Therefore, $\langle\xi\rangle$ and hence the LSSE signal is strongly suppressed by the application of high magnetic field.[20,108] However, the field induced suppression is dependent on the thickness of the magnetic film.[34] If the film thickness is lower than $\langle\xi\rangle$, the low frequency subthermal magnons cannot recognize the local temperature gradient and do not participate in LSSE. In that case, only high frequency magnons with shorter $\langle\xi\rangle$ and much higher energy than the Zeeman energy contribute towards the LSSE signal and hence the field induced suppression of LSSE becomes negligible.[20,108] However, if the film thickness is higher than $\langle\xi\rangle$, most of the low frequency magnons contribute towards LSSE and hence, the field induced suppression becomes more significant.[20,108] As we have observed in our TmIG films that the trend of temperature dependent LSSE signal is nearly independent of the substrate choice, the temperature dependent $\langle\xi\rangle$ is not supposed to change significantly with the substrate choice.

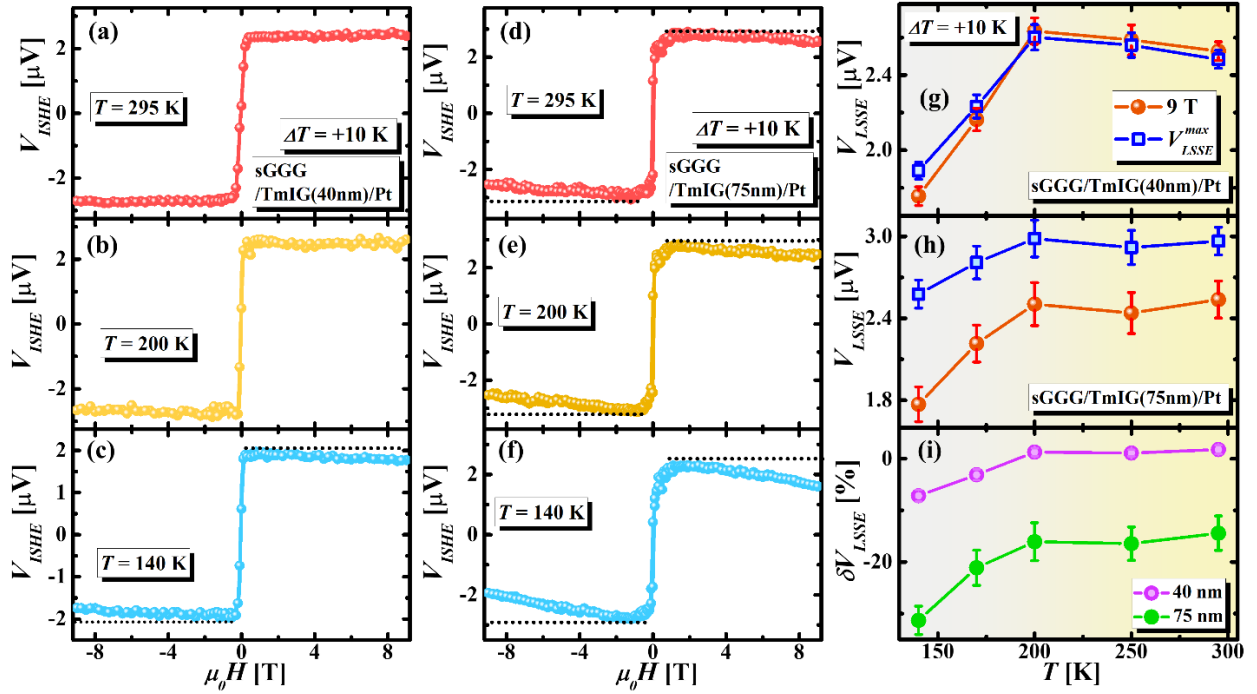

**Figure 9. Magnetic field induced suppression of the LSSE voltage:** $V_{ISHE}(H)$ loops for the sGGG/TmIG(40nm)/Pt nm film at $T$ = (a) 295, (b) 200 and (c) 140 K measured up to high magnetic field of $\mu_0 H = 9$ T. $V_{ISHE}(H)$ loops for the sGGG/TmIG(75nm)/Pt nm film at $T$ = (d) 295, (e) 200 and (f) 140 K measured up to high magnetic field of $\mu_0 H = 9$ T. $V_{LSSE}(T, \mu_0 H = 9\text{T})$ and $V_{LSSE}^{max}(T)$ for (g) 40 nm and (h) 75 nm films. (i) Temperature dependence of $\delta V_{LSSE}(\%)$ for 40 nm and 75 nm films.

Therefore, (i) to verify the influence of thickness on the field induced suppression of the LSSE signal in TmIG films and (ii) to confirm whether $\langle \xi \rangle$ of the GSGG/TmIG/Pt films obtained by analyzing the low field LSSE signal matches closely with that of the sGGG/TmIG/Pt films, we performed the high field LSSE measurements on the sGGG/TmIG/Pt films with thicknesses of 40 and 75 nm. **Figs. 9**(a)-(c) demonstrate the $V_{ISHE}(H)$ loops for the 40 nm film at $T$ = 295, 200 and 140 K measured up to high magnetic field of $\mu_0 H = 9$ T. It can be seen that the LSSE signal for the 40 nm film does not show prominent suppression at 9 T at 295 K. However, as temperature decreases below 200 K, the suppression of the LSSE signal becomes noticeable. On the other hand, as seen in **Figs. 9**(d)-(f), the LSSE signal for the 75 nm film shows significant suppression even at 295 K and the suppression of LSSE signal enhances with decreasing temperature. The more intense suppression of the LSSE signal in the 75 nm film compared to the 40 nm film at all temperatures between 295 and 140 K is also evident from $V_{LSSE}(T)$ for these two films shown in **Figs. 9**(g) and (h). We have also estimated the percentage change in $V_{LSSE}$ by the application of 9 T magnetic field, which we define as, $\delta V_{LSSE}(\%) = \left[\frac{V_{LSSE}(9\,\text{T}) - V_{LSSE}^{max}}{V_{LSSE}^{max}}\right] \times 100\%$, where, $V_{LSSE}(9\,\text{T})$ is the absolute value of $V_{LSSE}$ at 9 T magnetic field and $V_{LSSE}^{max}$ is the value of $V_{LSSE}$ at the maximum point of the $V_{ISHE}(H)$ loop. As shown in **Figs. 9**(i), $|\delta V_{LSSE}|$ for the 75 nm film is nearly 14% at 295 K but increases to ≈ 32% at 140 K. On the other hand, $|\delta V_{LSSE}|$ for the 40 nm film is negligible at 295 K but increases to ≈ 7% at 140 K. These results indicate that $\langle \xi \rangle$ for the sGGG/TmIG/Pt films at 295 K is between 40 and 75 nm, which is close to the value of $\langle \xi \rangle$ obtained for the GSGG/TmIG/Pt films. Since $\langle \xi \rangle$ decreases at low temperatures and becomes smaller than 40 nm below 150 K, the sGGG/TmIG(40nm)/Pt film shows significant field induced suppression of $V_{LSSE}$ at low temperatures. Similarly, since $\langle \xi \rangle$ at low temperatures is much smaller than 75 nm, the field induced suppression of $V_{LSSE}$ is also large at low temperatures for the 75 nm film. Note that in case of YIG/Pt films, the magnetic field induced suppression of the LSSE signal diminishes with decreasing temperature,[34] whereas, an opposite trend has been observed in case of TmIG. Such behavior can be explained by different trends of the temperature dependent $\langle \xi \rangle$ in YIG and TmIG. As explained by Guo *et al*.,[34] the temperature induced enhancement of $\langle \xi \rangle$ neutralizes the field induced suppression of $\langle \xi \rangle$, and because of these two competing factors, the field induced suppression of the LSSE voltage is less prominent at low temperatures in YIG/Pt films. On the contrary, the combined effects of the temperature induced reduction in $\langle \xi \rangle$ observed in our TmIG/Pt films and field induced suppression of $\langle\xi\rangle$ give rise to stronger field induced suppression of the LSSE voltage at lower temperatures.

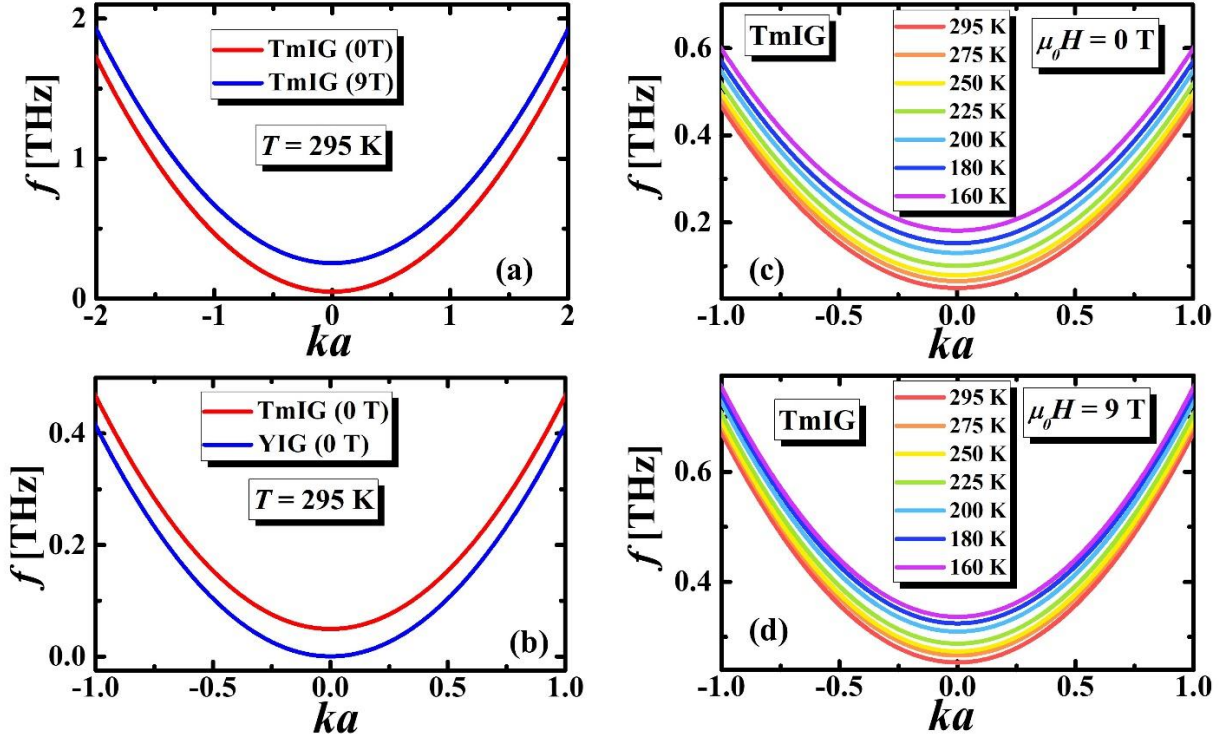

**Figure 10. Magnon frequency dispersion for TmIG:** (a) Magnon frequency dispersion for TmIG for $\mu_0 H = 0$ T and 9 T magnetic fields at $T = 295$ K. (b) Comparison of the magnon frequency dispersion for YIG and TmIG at room temperature for $\mu_0 H = 0$ T. Magnon frequency dispersion of TmIG at different temperatures for (c) $\mu_0 H = 0$ T and (d) 9 T magnetic fields.

Next, to have a qualitative understanding of the magnon frequency dependences of the LSSE signal and $\langle\xi\rangle$, we have estimated the magnon frequency dispersion for TmIG in the absence and in presence of high magnetic field of 9T. According to the classical Heisenberg ferromagnet model for spin waves, the parabolic magnon frequency dispersion at low energies can be expressed as,[17,20,109] $\hbar\omega_k = g_{eff}\mu_B H + D_{SW} \cdot k^2 a_0^2 + E_{ani}(K_{eff})$, where the first term represents the Zeeman energy gap due to the application of external magnetic field, the second term is associated with spin wave stiffness ($D_{SW}$ is the spin wave stiffness constant), and the third term represents the contribution of effective magnetic anisotropy energy, $E_{ani}(K_{eff})$. Here, the value of $D_{SW}$ is taken as that of YIG, *i.e.*, $D_{SW} a_0^2 = 4.2 \times 10^{-29}$ erg. cm² at room temperature.[108] Using the expression $K_{eff} = K_{shape} + K_{mc} + K_{me}$, we determined the temperature dependence of $E_{ani}(K_{eff})$ for TmIG. Here, the temperature variation of $K_{me}$ was obtained from the temperature dependence of $\lambda_{111}$ reported in the literature[83] (see **Supplementary Figures 8(e)**). Since the shear modulus, $c_{44}$ in REIGs is weakly dependent on the rare-earth species, the value of $c_{44}$ is taken as that of YIG (76.4 GPa at room temperature).[47] The temperature dependence of $K_{shape}$ was estimated from the temperature variation of $M_S$ (see **Supplementary Figures 8(f)**). We assumed constant value of $K_{mc} = 0.058$ kJ/m³ throughout the measured temperature range.[6] As shown in **Supplementary Figures 8(f)**, $K_{eff}$ is positive for $T \leq 300$ K ($K_{eff} = 20$ kJ/m³) and its absolute value increases considerably with decreasing temperature. **Fig. 10**(a) shows the magnon frequency dispersion for TmIG for $\mu_0 H = 0$ T and 9 T magnetic fields at $T = 295$ K. Clearly, the high magnetic field opens a magnon energy gap in the low frequency regime (much smaller than thermal energy at room temperature) indicating the suppression of $\langle \xi \rangle$ and hence significant reduction of the LSSE signal at high magnetic fields. As shown in **Fig. 10**(b), we have compared the magnon frequency dispersion for YIG and TmIG at room temperature for $\mu_0 H = 0$ T. It is evident that the opening of magnon energy gap in TmIG is higher than in YIG even in the absence of external magnetic field, which is mainly caused by the effective magnetic anisotropy. Note that $K_{eff}$ of TmIG is higher than that of YIG.[110] The higher value of magnon energy gap in TmIG compared to YIG thus indicates the higher possibility of freezing out of the low energy subthermal magnons in TmIG. This, along with higher value of $\alpha$ in TmIG contributes to the lower value of $\langle \xi \rangle$ and hence the LSSE voltage in TmIG compared to YIG[16]. Furthermore, the value of $g_{eff}$ at room temperature is lower in TmIG (≈1.63)[93] than in YIG (≈ 2.046)[111]. Therefore, for a given applied magnetic field strength, the magnitude of the magnon energy gap due to Zeeman effect will be different in TmIG than in YIG. Moreover, $\alpha$ in TmIG[6,96] is nearly two orders of magnitude higher than in YIG[112]. In other words, different values of $K_{eff}$, $g_{eff}$ and $\alpha$ as well as their different temperature dependences give rise to different temperature profiles of $\langle \xi \rangle$ in TmIG and YIG. In **Figs. 10**(c) and (d), we show the magnon frequency dispersion of TmIG at different temperatures for $\mu_0 H = 0$ T and 9 T magnetic fields, respectively. Clearly, the magnon energy gap increases with decreasing temperature due to enhanced magnetic anisotropy at low temperatures. Application of 9 T magnetic field increases the magnon energy gap further due to Zeeman effect. These results help explain the observed decrease in $\langle \xi \rangle$ and enhanced magnetic field induced suppression of the LSSE signal at low temperatures in TmIG.[20,108]

We believe that our findings will attract the attention of the spintronic community for further exploration of long-range thermo-spin transport in different REIG based magnetic thin films and heterostructures for tunable spincaloritronic efficiency by manipulating $H_K^{eff}$ and $\alpha$. For example, $H_K^{eff}$ of the REIG thin films grown on piezoelectric substrates can be modulated by applying a gate voltage,[113] which can eventually influence $\langle \xi \rangle$ and hence the spincaloritronic efficiency. Therefore, our study also provides a step towards the development of efficient spincaloritronic devices based on voltage controlled LSSE.

## 3. CONCLUSION

In summary, we have performed a comprehensive investigation of the temperature dependent LSSE, RF transverse susceptibility, and broadband FMR measurements on TmIG/Pt bilayers grown on different substrates. The decrease in the LSSE voltage below 200 K independent of TmIG film thickness and substrate choice is attributed to the increases in $H_K^{eff}$ and $\alpha$ that occur within the same temperature range. From the TmIG thickness dependence of the LSSE voltage, we determined the temperature dependence of $\langle\xi\rangle$ and highlighted its correlation with the temperature dependent $H_K^{eff}$ and $\alpha$ in TmIG/Pt bilayers, which will be beneficial for the development of REIG-based spincaloritronic nanodevices. Furthermore, the enhanced suppression of the LSSE voltage by the application of high magnetic field at low temperatures together with the temperature evolution of magnon frequency dispersion in TmIG estimated from the temperature dependent $K_{eff}$ and $\alpha$ support our observation of the decrement of $\langle\xi\rangle$ at low temperatures in the TmIG/Pt bilayers.

## 4. METHODS

*Thin film growth and structural/morphological characterization:* Single-crystalline TmIG thin films were deposited by pulsed laser deposition (PLD), using two different PLD setups. The thin films were grown epitaxially on different (111)-oriented substrates, including GGG ($Gd_3Ga_5O_{12}$), GSGG ($Gd_3Sc_2Ga_3O_{12}$), and sGGG (($Gd_{2.6}Ca_{0.4}$)($Ga_{4.1}Mg_{0.25}Zr_{0.65}$)$O_{12}$). Substrates with (111) orientation are chosen so that the magnetoelastic anisotropy of the TmIG films favors PMA. Using the first PLD setup, films with varying thickness between 28 nm and 236 nm were grown on GGG and GSGG substrates. A KrF excimer laser with a wavelength of 248 nm, a fluence of 3-4 J/cm², and a repetition rate of 2 Hz is used. Before the first deposition, the TmIG target was preablated inside the PLD chamber with more than $10^4$ pulses. All substrates were annealed for 8 h at 1250°C in oxygen atmosphere prior to the film deposition to provide a high substrate surface quality. Growth conditions were selected to achieve stoichiometric, single-crystalline thin films with a smooth surface of about 0.2-0.3 nm in root-mean square roughness (RMS). For all films, the substrate was heated to 595°C during the film deposition, monitored by a thermocouple inside the substrate holder. The TmIG thin films were grown at a rate of $0.01 - 0.02$ nm/s, in the presence of an oxygen background atmosphere of 0.05 mbar. After the deposition, the samples were cooled to room temperature at approximately 5 K/min, maintaining the oxygen atmosphere. A layer of 5 nm Pt was deposited at room temperature *ex-situ* on the garnet films by DC magnetron sputtering using a shadow mask. The TmIG films were annealed at 400°C for 1 h inside the sputter chamber prior to the Pt deposition to avoid surface contamination[114]. To complement these samples, TmIG films with thicknesses 75 and 40 nm were grown on sGGG substrates using a second PLD setup. The laser wavelength was 248 nm at 10 Hz, the fluence 1.3 J/cm², and the substrate temperature was ~750 °C with an oxygen pressure of 0.2 mbar. Samples were cooled at 20 K/min in 0.2 mbar oxygen.

The film surface morphology was investigated by atomic force microscopy (AFM), while the structural properties of the thin films were identified by x-ray diffraction (XRD) using monochromatic Cu Kα radiation. The film thickness was evaluated from the Laue oscillations (for the thinner films) and by spectroscopic ellipsometry. Further, a cross-sectional high resolution scanning transmission electron microscopy (HR-STEM) was conducted, using a JEOL NEOARM F200 operated at an electron energy of 200 keV. Electron energy loss spectra (EELS) were obtained using a GATAN Continuum S EELS spectrometer. The cross-sectional sample was prepared by mechanical dimpling and ion polishing. Interdiffusion between the TmIG film and substrate is expected to be limited to a depth of order 1-3 nm[41] and its effects are neglected for the film thicknesses used in this study. See **Supplementary Figure 1**(e) for energy dispersive X-ray spectroscopy (EDX) using transmission electron microscopy (TEM) performed on the GSGG/TmIG(205nm) film.

*Temperature dependent MFM measurements:* Temperature dependent MFM measurements were performed on a Hitachi 5300E system. All measurements were done under high vacuum (P ≤ $10^{-6}$ Torr). MFM measurements utilized HQ: NSC18/Co-Cr/Al BS tips, which were magnetized out-of-plane with respect to the tip surface via a permanent magnet. Films were first magnetized to their saturation magnetization by being placed in a 1T static magnetic field, in-plane with the film surface. After that AC demagnetization of the film was implemented before initiating the MFM scans. After scans were performed, a parabolic background was subtracted, which arises from the film not being completely flat on the sample stage. Then, line artifacts were subtracted before finally applying a small Gaussian averaging/sharpening filter over the whole image. Phase standard deviation was determined by fitting a Gaussian to the image phase distribution and extracting the standard deviation from the fit parameters.

*Magnetometry:* The magnetic properties of the samples were measured using a superconducting quantum interference device - vibrating sample magnetometer (SQUID-VSM) at temperatures between 10 K and 350 K. A linear background stemming from the paramagnetic substrate was thereby subtracted. Due to a trapped remanent field inside the superconducting coils, the measured magnetic field was corrected using a paramagnetic reference sample. Additionally, a polar magneto-optical Kerr effect (MOKE) setup was used to record out-of-plane hysteresis loops at room temperature. The molecular field coefficient (MFC) model was a Python-coded version of Dionne's model[115] using molecular field coefficients[57].

*Longitudinal spin Seebeck effect measurements:* The longitudinal spin Seebeck effect (LSSE) was measured over a broad temperature window of 120 K ≤ $T$ ≤ 295 K using a custom-built setup assembled on a universal PPMS sample puck. During the LSSE measurements, the films were sandwiched between two copper blocks, as shown in **Fig. 3**(a). The same sample geometry was used for all films and the distance between the contact leads on the Pt surface were fixed at $L_y$ = 3 mm for all films. A single layer of thin Kapton tape was thermally affixed to the naked surfaces of the top (cold) and bottom (hot) copper blocks. To ensure a good thermal link between the film surface and the Kapton tape (thermally conducting and electrically insulating) attached to the top and bottom blocks, cryogenic Apiezon N-grease was used. Additionally, the Kapton tape electrically insulated the cold (hot) blocks from the top (bottom) surface of the films. The temperatures of both these blocks were controlled individually by two separate temperature controllers (Scientific Instruments Model no. 9700) to achieve an ultra-stable temperature difference ($\Delta T$) with $[\Delta T]_{Error} < \pm 2$ mK. The top block (cold) was thermally anchored to the base of the PPMS puck using two molybdenum screws whereas a 4-mm-thick Teflon block was sandwiched between the puck base and the hot block (bottom) to maintain a temperature difference of ~ 10 K between the hot block and the PPMS base. A resistive chip-heater (PT-100 RTD sensor) and a calibrated Si-diode thermometer (DT-621-HR silicon diode sensor) were attached to each of these blocks to efficiently control and sense the temperature. The heaters and thermometers attached to the copper blocks were connected to the temperature controllers in such a manner that a temperature gradient develops along the +$z$-direction that generates a temperature difference, $\Delta T$, between the top (cold) and bottom (hot) copper blocks. For a given temperature gradient, the in-plane voltage generated along the $y$-direction across the Pt layer due to the ISHE ($V_{ISHE}$) was recorded by a Keithley 2182a nanovoltmeter while sweeping an external in-plane DC magnetic field from positive to negative values along the $x$-direction. The Ohmic contacts for the voltage measurements were made by electrically anchoring a pair of ultra-thin gold wires (25 µm diameter) to the Pt layer by high quality conducting silver paint (SPI Supplies).

*Transverse susceptibility measurements:* The temperature evolution of effective magnetic anisotropy in the GSGG/TmIG/Pt film was measured by employing a radio frequency (RF) transverse susceptibility (TS) technique using a home-built self-resonant tunnel diode oscillator (TDO) circuit with a resonance frequency of 12 MHz and sensitivity of ±10 Hz. A physical property measurement system (PPMS) was employed as a platform to scan the external DC magnetic field ($H_{DC}$) and temperature. Before the TS measurements, the film was mounted inside an inductor coil (L), which is a component of an LC tank circuit. The entire tank circuit was placed outside the PPMS except the coil, L, which was positioned at the base of the PPMS sample chamber using a multi-purpose PPMS probe inserted in such a manner that the axial RF magnetic field ($H_{RF}$) of amplitude ~ 10 Oe produced inside the coil was always parallel to the film surface, but perpendicular to $H_{DC}$. For the TmIG with IP easy axis, $H_{DC} \perp$ film surface, whereas for the films with OOP easy axis, $H_{DC} \parallel$ film surface. When the sample is subject to both $H_{RF}$ and $H_{DC}$, the dynamic susceptibility of the sample changes which in turn changes the inductance of the coil and, hence, the resonance frequency of the LC tank circuit. The relative change in the resonance frequency is proportional to the relative change in the transverse susceptibility of the sample. Therefore, TS as a function of $H_{DC}$ was acquired by monitoring the shift in the resonance frequency of the TDO-oscillator circuit by employing an Agilent frequency counter.

*Broadband ferromagnetic resonance measurements:* Broadband ferromagnetic resonance (FMR) measurements ($f$ = 6-20 GHz) were performed using a broadband FMR spectrometer (NanOsc[TM] Phase-FMR Spectrometer, Quantum Design Inc., USA) integrated to a Dynacool PPMS. The TmIG film was firmly affixed on the surface of a commercial 200-μm-wide coplanar waveguide (CPW) (also provided by NanOsc[TM] Phase-FMR Spectrometer, Quantum Design Inc., USA) using Kapton tape. The TmIG films were placed faced down on the CPW so that the CPW can efficiently transmit the MW signal from the RF source over a broad *f*-range. The role of the Kapton tape is to electrically insulate the films from the CPW. An in-plane RF magnetic field, $H_{RF}$ is generated in close vicinity to the CPW. In presence of an appropriate external in-plane DC magnetic field, $H_{DC}$ provided by the superconducting magnet of the PPMS applied along the direction of the MW current flowing through the CPW ($H_{DC} \perp H_{RF}$) and frequency, $H_{RF}$ resonantly excites the TmIG film. The spectrometer employs lock-in detection and records the field derivative of the power absorbed ($dP/dH$) by the film when it is excited by a microwave (MW) electromagnetic field generated by injecting a MW current to the CPW.

## ACKNOWLEDGEMENTS

Financial support by the US Department of Energy, Office of Basic Energy Sciences, Division of Materials Science and Engineering under Award No. DE-FG02-07ER46438 at USF and by the German Research Foundation (DFG) within project No. 318592081AL618/37-1 at U Augsburg are gratefully acknowledged. CR acknowledges support of NSF award DMR 1808190 and 1954606.

## CONFLICT OF INTEREST

The authors have no conflicts to disclose.

## DATA AVAILABILITY

The data that support the findings of this study are available from the corresponding author upon reasonable request.